\documentclass[journal]{IEEEtran}
\usepackage{cite}
\usepackage[pdftex]{graphicx}
\usepackage[cmex10]{mathtools}
\usepackage{amssymb}
\usepackage{epsfig}
\usepackage{subfigure}
\usepackage{algorithm}
\usepackage{algpseudocode}
\usepackage[usenames,dvipsnames]{color}
\usepackage{array}

\usepackage{bm}
\usepackage{bbm}
\usepackage{microtype}
\usepackage{kotex}

\newtheorem{lemma}{Lemma}

\newtheorem{remark}{Remark}
\newtheorem{assumption}{Assumption}

\def\cA{{\mathcal{A}}}   
 \def\cF{{\mathcal{F}}}

\def\ba{{\mathbf{a}}}  \def\bc{{\mathbf{c}}} \def\bd{{\mathbf{d}}}
\def\bee{{\mathbf{e}}} \def\bff{{\mathbf{f}}}  \def\bh{{\mathbf{h}}}
  \def\bk{{\mathbf{k}}} 
   
 \def\br{{\mathbf{r}}}  
\def\bu{{\mathbf{u}}} \def\bv{{\mathbf{v}}} \def\bw{{\mathbf{w}}} \def\bx{{\mathbf{x}}}
 \def\bz{{\mathbf{z}}}  

\def\bA{{\mathbf{A}}}  \def\bC{{\mathbf{C}}} \def\bD{{\mathbf{D}}}
 \def\bF{{\mathbf{F}}}  \def\bH{{\mathbf{H}}}

 \def\bR{{\mathbf{R}}}  
\def\bU{{\mathbf{U}}}

\DeclareMathOperator{\rank}{rank}
\DeclareMathOperator*{\argmin}{arg\,min}
\DeclareMathOperator*{\argmax}{arg\,max}

\begin{document}


\title{Advanced Quantizer Designs for FDD-Based FD-MIMO Systems Using Uniform Planar Arrays}


\author{Jiho~Song,~\IEEEmembership{Member,~IEEE,} Junil~Choi,~\IEEEmembership{Member,~IEEE,} {Taeyoung~Kim},~\IEEEmembership{Member,~IEEE,} and David~J.~Love,~\IEEEmembership{Fellow,~IEEE}
\thanks{J.\ Song is with School of Electrical Engineering, University of Ulsan, Ulsan, 44610, Korea (e-mail:jihosong@ulsan.ac.kr).}
\thanks{J.\ Choi is with the Department of Electrical Engineering, POSTECH, Pohang, Gyeongbuk 37673, Korea (e-mail:junil@postech.ac.kr).}
\thanks{T.\ Kim, is with Samsung Electronics Co., Ltd., Suwon, Korea (email:ty33@samsung.com).}
\thanks{D.\ J.\ Love is with the School of Electrical and Computer Engineering, Purdue University, West Lafayette, IN 47907 (e-mail:djlove@purdue.edu).}
\thanks{Parts of this paper were presented at the Globecom, Washington, DC USA, December 4-8, 2016 \cite{Ref_Son16_con}.}
\thanks{This work was supported in part by Communications Research Team (CRT), Samsung Electronics Co. Ltd., and the ICT R$\&$D program of MSIP/IITP  [2017(B0717-17-0002), Development of Integer-Forcing MIMO Transceivers for 5G $\&$ Beyond Mobile Communication Systems].}
}

\maketitle


\begin{abstract}
Massive multiple-input multiple-output (MIMO) systems, which utilize a large number of antennas at the base station, are expected to enhance network throughput by enabling improved multiuser MIMO techniques. To deploy many antennas  in reasonable form factors,  base stations are expected to employ antenna arrays in both horizontal and vertical dimensions, which is known as full-dimension (FD) MIMO. The most {popular} two-dimensional array is the uniform planar array (UPA), where antennas are placed in a grid pattern. To exploit the full benefit of massive MIMO in frequency division duplexing (FDD), the downlink channel state information (CSI) should be estimated, quantized, and fed back from the receiver to the transmitter. However, it is difficult to accurately quantize the channel in a computationally efficient manner due to the high dimensionality of the massive MIMO channel. In this paper, we develop both narrowband and wideband CSI quantizers  for FD-MIMO taking the properties of realistic channels and the UPA into consideration. To improve quantization quality, we focus on not only quantizing dominant radio paths  in the channel, but also combining the quantized beams. We also develop a hierarchical beam search approach, which scans both vertical and horizontal domains jointly with moderate computational complexity. Numerical simulations verify that the performance of the proposed quantizers is better than that of previous CSI quantization techniques.
\end{abstract}


\section{Introduction}

\IEEEPARstart{M}{assive} multiple-input multiple-output (MIMO) systems {are} a strong candidate to fulfill the throughput requirements for fifth generation (5G) cellular networks \cite{Ref_Mar10}.  {To maximize the number of antennas in a limited area, two-dimensional  antenna arrays, e.g., uniform planar arrays (UPAs) and cylindrical arrays, that host antennas in both vertical and horizontal domains are being prominently considered in practice \cite{Ref_Han98,Ref_Nam13}. Among various 2D array solutions, UPAs are of great interest to simplify signal processing for three-dimensional (3D) channels for FD-MIMO.} Massive MIMO employing a UPA structure is known as full-dimension (FD) MIMO because of its ability to exploit both vertical and horizontal {domain beamforming} \cite{Ref_3GPP15381,Ref_3GPP150057,Ref_3GPP150560,Ref_Li13,Ref_Cho15,Ref_Cho15_con}.

{To fully exploit FD-MIMO, accurate channel state information (CSI) for both domains is critical.}  Channel estimation techniques relying upon time division duplexing (TDD) can leverage channel reciprocity if the transmit and receive arrays are calibrated \cite{Ref_Ngo13,Ref_Rog14}. {Moreover, it has been recently verified that the spectral efficiency of TDD-based massive MIMO increases without bound as the number of antennas grows, even under pilot contamination \cite{Ref_Bjo18}.} Most current cellular systems, however, exploit frequency division duplexing (FDD) where the receiver should estimate, quantize, and feed back the downlink CSI to the transmitter. The high dimensionality of a massive MIMO channel {could} cause {large overheads for downlink} channel training and quantization processes \cite{Ref_Has03,Ref_AuY07,Ref_AuY11,Ref_Cho13,Ref_Cho15JJ,Ref_Cho14,Ref_Noh14}. We focus on CSI quantization for FD-MIMO in this paper and refer to \cite{Ref_Cho14,Ref_Noh14,Ref_Han17} {and the references therein for the massive MIMO downlink training problem.}

The majority of CSI quantization codebooks have been designed under the assumption of spatially uncorrelated Rayleigh fading channels, which  are uniformly distributed on the unit hypersphere when normalized. {To quantize these channels, the codewords in a codebook should cover the unit} sphere as uniformly as possible \cite{Ref_Muk03,Ref_Lov03,Ref_Lov08}. For spatially correlated channels, the codebooks have been carefully shaped based on the prior knowledge of {channel statistics \cite{Ref_Lov04,Ref_Lov06,Ref_Xia06,Ref_Rag07,Ref_Rag15}.}

Although most previous codebooks have been designed based on analytical channel models, it is difficult to represent   the properties of true three-dimensional (3D) channels for FD-MIMO. {The 3D spatial channel model (SCM) in \cite{Ref_Nam13,Ref_3GPP14}, which is an extension of the 2D SCM \cite{Ref_3GPP03}, has been extensively used to mimic the measured channel variations {for 3GPP standardization.} Although the 3D SCM is a stochastic channel model, {it provides limited insights}  into  practical CSI quantizer designs. Therefore, it is necessary to} develop a simple channel model that accurately represents the properties of the 3D-SCM channel with UPA antennas.


In this paper, we define a simple 3D channel model {using} the sum of a finite number of scaled  array response vectors. {Based on this simplified channel model, we develop CSI quantizers for UPA scenarios. We first carry out performance analysis of Kronecker product (KP) CSI quantizers  \cite{Ref_Son16_con,Ref_Cho15}. Our analytical studies on the KP CSI quantizers provide design guidelines on how to develop a quantizer for a narrowband, single frequency tone CSI} using  limited feedback resources. In the proposed quantizer, we concentrate on detecting/quantizing dominant radio paths in true channels. {To maximize quantization quality, we also develop a codebook for combiners that cophases and scales the quantized beams.} Both vertical and horizontal domains are {searched during beam quantization, which involves a heavy computational complexity.} We  thus develop a hierarchical beam search approach  to reduce the complexity.

We also develop a wideband quantizer for broadband communication by evolving  the dual codebook structure in LTE-Advanced \cite{Ref_3GPP103378,Ref_3GPP105011}. In the dual codebook structure,  a first layer quantizer is used to search correlated CSI between multiple frequency tones. Unless dominant paths are gathered in a single cluster, the LTE-Advanced codebook is not effective because only adjacent radio paths are selected and quantized  using the same resolution  codebook.  We thus concentrate on detecting adjacent and/or sperate paths within each  wideband resource block (RB) based on the proposed hierarchical beam search approach. In addition, {a} second layer quantizer is designed to refine the beam direction of the quantized wideband CSI according to the channel vectors in a narrowband RB. When comparing our approach to the LTE-Advanced codebook, the refined beams are cophased and scaled in our approach, while the LTE-Advanced codebook {only cophase} adjacent beams without considering beam refinement.

In Section II, we describe FD-MIMO systems employing UPAs and discuss a simple channel model that mimics true {3D SCM channels.} In Section III, we review previously reported KP codebooks. {In Section IV, we develop {a narrowband CSI quantizer that takes multiple radio paths into account and conduct performance analysis to develop a design guideline for CSI quantizers. In Section V, we also propose a wideband CSI quantizer assuming a multi-carrier framework.} In Section VI, we present simulation results, and the conclusion follows in Section VII.

Throughout this paper, $\mathbb{C}$ denotes the field of complex numbers, {$\mathbb{N}$ denotes the semiring of natural numbers,} {$\mathcal{CN}(\mu,\sigma^2)$} denotes the complex normal distribution with mean $\mu$ and variance $\sigma^2$, $[a,b]$ is the closed interval between $a$ and $b$, $\mathrm{U}(a,b)$ denotes the uniform distribution in the closed interval $[a,b]$, $\lceil~\rceil$ is the ceiling function, {$\Gamma(n)$ denotes the complete gamma function,} {$\mathrm{E}_{\mathrm{a}}[\cdot]$ is the expectation of independent random variable $\mathrm{a}$}, $\| \cdot \|_p$ is the $p$-norm, $\odot$ is the Hadamard product, $\otimes$ is the Kronecker product, {$\mathbf{0}_{a}$ is the $a \times 1$ all zeros vector, $\mathbf{I}_{N}$ is the $N \times N$ identity matrix, {$\mathbf{1}_{a,b}$ is the $a \times b$ all ones matrix, $\bA_{a,b}$, {$\mathfrak{v}_{n}\{\bA\}$, and $\mathfrak{eig}_{n}\{\bA\}$  denote $(a,b)^{th}$ entry, the $n$-th dominant eigenvector, and the {$n$-th dominant eigenvalue} of the matrix $\bA$.}} Also, { $\ba^H$, $\ba^*$, $(\ba)_{\ell}$, $\ba_{[a:b]}$ denote the conjugate transpose,  element-wise complex conjugate, $\ell$-th entry, and subvector including entries between $[a,b]$ of the column vector $\ba$, respectively.

\section{System Model}
We consider multiple-input single-output (MISO) systems\footnote{{Although we mainly discuss MISO channel quantization to simplify presentation, the proposed channel quantizer can be easily extended to multiple-input multiple-output (MIMO) systems. The extension of the MISO channel quantizer will be discussed in Section \ref{sec_J4:mu_mimo}.}} employing $M \doteq M_vM_h$ {transmit antennas at the base station and a single receive antenna at the user, where $M_v$ is the number of rows and $M_h$ is the number of columns of the UPA antenna structure \cite{Ref_Han98}.} 
Assuming a multi-carrier framework, an input-output expression  is defined as
\begin{align}
\label{eq_J4:01}
 {y[w]=\sqrt{\rho}\bh[w]^{H}\bff[w] s[w] + n[w],}
\end{align}
where $y[w]$ is the received baseband symbol, $\rho$ is the signal-to-noise ratio (SNR),  $\bh[w] \in \mathbb{C}^M$ is the block fading MISO channel, {$\bff[w] \in \mathbb{C}^M$ is the {unit-norm} transmit beamformer,} $s[w] \in \mathbb{C}$ is the data symbol with the power constraint $\mathrm{E}[|s[w]|^2]\leq 1$, and $n[w] \sim \mathcal{CN}(0,1)$ is the additive white Gaussian noise. Note that $w \in \{1,\cdots, W \}$ {denotes the} frequency tone in the multi-carrier framework.



{To facilitate quantizer designs, we mimic the 3D-SCM and define a {simplified} channel model with a few radio paths  according to}
\begin{align}
\label{eq_J4:channel_finite}
&{\bh}[w]   \doteq \sum_{p=1}^{P}    e^{-j 2\pi \big(w-\frac{W-1}{2}\big) \triangle t_{p}} \alpha_p\mathbf{d}_{M}(\psi_{p}^{v},\psi_{p}^{h},w),
\end{align}
{while the true 3D-SCM is used to present numerical results in Section \ref{sec_J4:Sim}. In (\ref{eq_J4:channel_finite}),} $P$ is the number of dominant paths, $\triangle$ is the subcarrier spacing, $t_p$ is the excess tap delay of the $p$-th radio path, $\alpha_p $ is the channel gain of the $p$-th radio path, and $\mathbf{d}_{M}(\psi_{p}^{v},\psi_{p}^{h},w)$ is the $p$-th radio path {at given angles $(\psi_{p}^{v},\psi_{p}^{h})$.} In the UPA scenario, a radio path  for the $w$-th frequency tone is represented as
\begin{align}
\label{eq_J4:ray}
\mathbf{d}_{M}(\psi^{v},\psi^{h},w) \doteq \mathbf{d}_{M_v}(\psi^{v},w) \otimes \mathbf{d}_{M_h}(\psi^{h},w)
\end{align}
where the array response vector $\mathbf{d}_{M_a}(\psi_{a},w)$ is defined as
\begin{align}
\label{eq_J4:ar}
\mathbf{d}_{M_a}(\psi^a,w) \doteq \frac{1}{\sqrt{M_a}} \big[ 1,~e^{j\frac{2\pi d_a}{\lambda[w]}\psi^a},\cdots,e^{j\frac{2\pi d_a}{\lambda[w]} ({M_a}-1)\psi^a} \big]^{T},
\end{align}
for $a \in\{v,h\}$. In (\ref{eq_J4:ar}),  $\psi^v= \sin \phi^v$ and  $\psi^h= \sin \phi^h \cos \phi^v$ \cite{Ref_Han14}. {Note that  $d_a$ is the antennas spacing, $\phi^a$ is the angle for the array  vector,} and $\lambda[w]$ is the wavelength for the {$w$-th frequency tone CSI}
\begin{align}
\lambda[w]\doteq \frac{\lambda_c}{1+\frac{\triangle}{f_c}\big(w-\frac{W+1}{2}\big)}
\end{align}
where $f_c$ is the center frequency satisfying $c=f_c \lambda_c$ with the speed of light $c$. Without loss of generality, {a narrowband representation of  channels is} defined by plugging\footnote{{We consider the $\frac{W+1}{2}$-th subcarrier to ignore beam squinting effects \cite{Ref_Mai05}.}} $w=\frac{W+1}{2}$ into (\ref{eq_J4:channel_finite}) 
\begin{align}
\label{eq_J4:simple_channel}
{\bh} &= \sum_{p=1}^{P} \mathbf{d}_{M}(\psi_{p}^{v},\psi_{p}^{h}) \alpha_p
\\
\nonumber
&=  \bD \ba
\end{align}
where $\bD \doteq \big[\bd_{M}(\psi_{1}^v,\psi_{1}^h),\cdots,\bd_{M}(\psi_{p}^{v},\psi_{p}^{h})\big] \in \mathbb{C}^{M \times P}$ is the set of radio paths and $\ba \doteq [\alpha_1,\cdots,\alpha_P]^T \in \mathbb{C}^{P}$ is the set of complex channel gains. {In the narrowband assumption, $w$  is  dropped for simplicity.} We assume that {the beam directions are uniformly distributed in both vertical and horizontal domains such as   $\psi_{p}^{v},\psi_{p}^{h} \sim \mathrm{U}(-1,1)$ and are independent of channel gains $\alpha_p \sim \mathcal{CN}(0,1)$.}

{In the limited feedback beamforming approach, each user chooses a transmit beamformer  among codewords in the codebook $\cF \doteq \{ \bff_1,\cdots, \bff_{2^{B_{\textrm{T}}}}\}$ such that
\begin{align*}
\bff=\argmax_{\tilde{\bff} \in \cF} |\bh^H\tilde{\bff}|^2,
\end{align*}
where $B_{\textrm{T}}$ denotes the total feedback overhead. Based on the assumption that both transmitter and receiver know the predefined codebook, the $B_{\textrm{T}}$-bit index of the selected beamformer is fed back to the transmitter over the feedback link.}

{The majority of channel quantization codebooks have been designed for {spatially correlated and uncorrelated Rayleigh fading channels \cite{Ref_Muk03,Ref_Lov03,Ref_Lov08,Ref_Lov04,Ref_Lov06,Ref_Xia06,Ref_Rag07,Ref_Rag15}.  These} analytical channel models rely upon rich scattering environments so that each radio path has a limited effect on channel characterization. Thus, {most previous beamformer codebooks} focus on covering the unit hypersphere  as uniformly as possible without considering each radio path individually. However, the analytical channel models  are much different than realistic channel models that  assume only a few dominant scatterers. {Therefore}, this {line-packing} codebook design approach may not be effective when the number of antennas is large. To accurately quantize high-dimensional massive MIMO channels, {it is important to} tailor the codebook to the realistic channels consisting of a limited number of radio paths.}

\section{Kronecker Product Codebook Review - Single Beam Case}
\label{sec_J4:review}

It is critical {in FDD} massive MIMO systems to quantize and feedback information about the high-dimensional channels to the transmitter \cite{Ref_Has03,Ref_AuY07,Ref_AuY11,Ref_Cho13,Ref_Cho15JJ,Ref_Cho14,Ref_Noh14}. Thus, CSI quantization codebooks have been developed to tailor the feedback link with limited overhead  to the massive MIMO channels \cite{Ref_AuY11,Ref_Cho13,Ref_Cho15JJ}. Among various CSI quantization techniques,\footnote{{In massive FD-MIMO systems, formulating a low-dimensional feature abstraction problem for quantizing true channel vectors would be an interesting topic for future research.}} KP codebooks are of great interest to quantize the channels in a computationally efficient manner by considering the 2D antenna structure \cite{Ref_Cho15,Ref_Cho15_con}. Based on the  channel model in (\ref{eq_J4:simple_channel}), KP codebooks {are designed to quantize a} radio path with UPA structure \cite{Ref_3GPP150057,Ref_3GPP150560,Ref_Li13}.

 Most KP codebooks are based on the assumption that the covariance matrix of the channels is approximated by the KP of covariance matrices of  vertical and horizontal domains  such that \cite{Ref_Yin14}
\begin{align*}
\bR_{\bh} &\doteq \mathrm{E}_{\bh}\big[ \bh \bh^H\big]
\\
&\simeq \bR^{v} \otimes \bR^{  h}.
\end{align*}
Thus, a KP codebook is of the form
\begin{align*}
\cF = \big\{ {\bff} \in \mathbb{C}^M :   {\bff} = \bc^v \otimes (\bc^h)^*,~\bc^v \in \cA_{B_{\textrm{T}}/2}^v,~\bc^h \in \cA_{B_{\textrm{T}}/2}^h \big\},
\end{align*}
{to quantizes a}  single dominant path with the discrete Fourier transform (DFT) codebook
\begin{align}
\label{eq_J4:DFTCB}
\mathcal{A}_{B}^{a} = \big\{ \ba_{M_a}({1}/{Q}),\cdots, \ba_{M_a}({Q}/{Q}) \big\}
\end{align}
consisting of the $Q=2^{B}$ codewords
\begin{align*}
\ba_{M_a}(q/Q) \doteq \frac{1}{\sqrt{M_a}} \Big[1,~e^{j\pi(\frac{2 q}{Q}-1)}, \cdots, e^{j\pi(M_a-1)(\frac{2  q}{Q}-1)} \Big]^{T}.
\end{align*}

Most previous KP codebooks quantize the first dominant vector in each domain separately \cite{Ref_3GPP150057,Ref_3GPP150560,Ref_Li13,Ref_Cho15}. To find singular vectors, the channel vector $\bh$  in (\ref{eq_J4:simple_channel}) is decomposed  into both vertical and horizontal domains based on the singular value decomposition \cite{Ref_Cho15} yielding
\begin{align}
\label{eq_J4:rewritten}
\bh & = \sum_{k=1}^{\rank{({\bar{\bH}})}} (\bu_k \otimes \bv_k^*)\sigma_k,
\end{align}
where the reshaped channel is given in a matrix form
\begin{align*}
\bar{\bH} \doteq \big[\bh_{[1:M_h]},\cdots,\bh_{[(M_v-1)M_h+1:M_vM_h]} \big]^T  \in \mathbb{C}^{M_v \times M_h}.
\end{align*}
In (\ref{eq_J4:rewritten}), $\sigma_k$ denotes the $k$-th dominant singular value, $\bu_k \in \mathbb{C}^{M_v}$ denotes the $k$-th dominant left singular vector, and $\bv_k \in \mathbb{C}^{M_h}$ denotes the $k$-th dominant right singular vector of $\bar{\bH}$. The final codeword is then  obtained as $\bff= {\bc}^{v}\otimes ({\bc}^{h})^*$, where
\begin{align*}
{\bc}^{v}=\argmax_{\tilde{\bc}^{v} \in \mathcal{A}_{{B_{\textrm{T}}/2}}^{v}}{|\bu_1^H\tilde{\bc}^{v}|^2},~~~{\bc}^{h}=\argmax_{\tilde{\bc}^{h} \in \mathcal{A}_{{B_{\textrm{T}}/2}}^{h}}{|\bv_1^H\tilde{\bc}^{h}|^2}.
\end{align*}

{Despite the advantage of a KP codebook, it has some issues. Even with a line-of-sight (LOS) channel,} a dominant radio path may not be accurately quantized by searching each domain separately. Also, it is not {always} effective to quantize only a single radio path  because even $\bu_1 \otimes \bv_1^*$ may consist of multiple  paths.  Although the quantizer in \cite{Ref_Cho15} considers adding two beams in $\bu_1 \otimes \bv_1^*$, the performance improvement is limited because the beams are not combined properly.

\section{{Proposed Narrowband Quantizer - Multiple Beams Case}}

{Prior work has verified that most 3D SCM channel realizations are well modeled with only a few resolvable 2D radio paths \cite{Ref_Son16_con,Ref_Cho15}.  Thus, we assume that the channel vector on a single frequency tone can be represented by a combination of a set of multiple radio paths and its corresponding channel gain vector   \cite{Ref_Son16_con}.  In (\ref{eq_J4:simple_channel}), the array response vectors in $\bD$ and the channel gain vector $\ba$ contain different types of channel information. Thus, we focus on quantizing $\bD$ and $\ba$ using different codebooks in this paper.}

In our narrowband quantizer, we aim to find
\begin{align}
\label{eq_J4:mimic}
\frac{\bh}{\|\bh \|_2} &\simeq   \frac{\bC\bz}{\|\bC\bz \|_2}
\end{align}
by constructing a set of $N$ beams $\bC \doteq [ \bc_{1},\cdots,\bc_{N} ] \in \mathbb{C}^{M \times N}$ and {a {unit-norm}} weight vector $\bz \doteq [ z_{1},\cdots, z_{N} ]^T \in \mathbb{C}^{N}$. The radio paths constituting the channels are represented by the Kronecker product of array response vectors as in (\ref{eq_J4:ray}). Thus, each 2D beam in $\bC$ can be defined by a combination of quantized array response vectors in  vertical and horizontal domains such that
\begin{align*}
\bc_{n}\doteq \bc_{n}^{v} \otimes \bc_{n}^{h}, \quad n=1,\cdots,N.
\end{align*}
We assume that $B_n$-bit is reserved to quantize the $n$-th array response vector $\bc_{n}^{a}$ in the domain $a$ and $B_c$-bit is reserved to quantize the weight vector $\bz$.

To construct $\bC$ and $\bz$ under the condition of {a limited feedback overhead of $B_{\mathrm{T}}={B}_c+\sum_{n=1}^{{N}} 2{B}_{n}$-bits,} the following questions should be properly addressed.
\\
\textbf{1) Beam quantization:} How should the radio paths in the channel be chosen and quantized?
\\
\textbf{2) Beam combining:} How should the quantized beams be cophased and scaled?
\\
\textbf{3) Feedback resource allocation:}  {For a given total feedback overhead $B_{\mathrm{T}}$, how should the feedback-bit allocation scenario
\begin{align}
\label{eq_J4:fba}
\br_N \doteq [B_1,\cdots,B_N,B_c]^T \in \mathbb{N}^{{N}+1}
\end{align}
be defined to effectively quantize and combine the limited number of radio paths?}
\\
{In the following subsections, we address our channel quantization procedure across two separate quantization phases and evaluate quantization loss at each phase.}
\begin{remark}
\label{rm_J4:01}
{The quantized channel vector can be viewed as a representation of the channel using} an analog beamsteering matrix $\bC$, which is realized by a set of radio frequency phase shifters, and a baseband beamformer $\bz$. {Therefore, the proposed approach follows the hybrid beamforming architecture in \cite{Ref_Aya14}.}
\end{remark}

\subsection{Phase I: Beam Quantization}
\label{sec_J4:beam quantization}

\begin{figure*}[!t]
\setcounter{equation}{18}
\begin{align}
\label{eq_J4:cov}
\bR & = \left[\begin{array}{ccc} \mathrm{E}_{\bh} [  ( \bc_{1}^{v} \otimes  \bc_{1}^{h})^H {{\bh}} {{\bh}}^H (\bc_{1}^{v} \otimes  \bc_{1}^{h}) ]&  \cdots & \mathrm{E}_{\bh} [(\bc_{1}^{v} \otimes  \bc_{1}^{h})^H {{\bh}} {{\bh}}^H (\bc_{N}^{v}\otimes  \bc_{N}^{h}) ] \\
\vdots &    \ddots  &     \vdots
 \\
\mathrm{E}_{\bh} [(\bc_{N}^{v}\otimes  \bc_{N}^{h})^H {{\bh}} {{\bh}}^H (\bc_{1}^{v} \otimes  \bc_{1}^{h}) ] &  \cdots   &      \mathrm{E}_{\bh} [(\bc_{N}^{v}\otimes  \bc_{N}^{h})^H {{\bh}} {{\bh}}^H (\bc_{N}^{v}\otimes  \bc_{N}^{h})
 \\
\end{array} \right].
\end{align}
\setcounter{equation}{10}
\hrulefill
\end{figure*}

{In the beam quantization phase, we aim to construct a selected set of quantized 2D beams $\bC$.}  {It is well known that the DFT codebook is an effective solution to quantize array response vectors so that we quantize the $n$-th array response vector $\bc_n^a$ with the DFT codebook $\cA_{B_n}^a$.}

{In our beam quantization approach,  we select and quantize the 2D DFT beams sequentially.} In the $n$-th update, the quantized DFT vectors and the unquantized weight vector\footnote{The bar on the top of weight vectors {denotes the unquantized weight vectors.}} are obtained by solving the maximization problem\footnote{{Assuming a multiuser framework,  the interuser interference can be suppressed by maximizing the beamforming gain, i.e., minimizing the quantization error.}}
\begin{align}
\label{eq_J4:OPBQ}
({\bc}_{n}^{v},{\bc}_{n}^{h}, \bar{\bz}_n)=\argmax_{(\tilde{\bc}^{v},\tilde{\bc}^{h},\tilde{\bz} ) \in \mathcal{A}_{B_n}^{v} \times \mathcal{A}_{B_n}^{h}  \times \mathbb{C}^{n}}{ \frac{\big|{\bh}^H\bC_{n|\tilde{\bc}^{v},\tilde{\bc}^{h}} \tilde{\bz} \big|^2}{\big\|\bC_{n|\tilde{\bc}^{v},\tilde{\bc}^{h}}\tilde{\bz}  \big\|_2^2} },
\end{align}
where $\bC_{n|\tilde{\bc}^{v},\tilde{\bc}^{h}} \doteq [ \bc_1,\cdots, \bc_{n-1} ,\tilde{\bc}^{v} \otimes \tilde{\bc}^{h} ] \in \mathbb{C}^{M \times n}$, and $\{  \bc_1,\cdots, \bc_{n-1}\}$ includes previously selected DFT beams.

\begin{algorithm}
  \caption{Beam quantization}
  \label{Al_J4:01}
  \begin{algorithmic}
\State \textbf{Initialization}
\State 1:~{Create an initial empty matrix ${\bC}_{0}$}
{\State \textbf{Beam quantization}}
\State 2:~\textbf{for}~$1 \leq n \leq  N$
{\State 3:~~Given DFT codebooks $(\tilde{\bc}^{v},\tilde{\bc}^{h}) \in \mathcal{A}_{B_n}^{v} \times \mathcal{A}_{B_n}^{h}$}
{\State 4:~~Construct beam set $\bC_{n|\tilde{\bc}^{v},\tilde{\bc}^{h}}=[ \bC_{n-1}, \tilde{\bc}^{v} \otimes \tilde{\bc}^{h} ]$}
{\State 5:~~Quantize radio path $\bc_n = \bc_{n}^{v} \otimes \bc_{n}^{h}$, where}
{\State ~~~~$(\bc_{n}^{v},\bc_{n}^{h})=\argmax_{\tilde{\bc}^{v},\tilde{\bc}^{h} \in \mathcal{A}_{B_n}^{v} \times \mathcal{A}_{B_n}^{h}}$}
{\State ~~~~~~$\mathfrak{eig}_{1} \big\{(\bC_{n|\tilde{\bc}^{v},\tilde{\bc}^{h}}^H \bC_{n|\tilde{\bc}^{v},\tilde{\bc}^{h}})^{-1}\bC_{n|\tilde{\bc}^{v},\tilde{\bc}^{h}}^H{\bh}{\bh}^H \bC_{n|\tilde{\bc}^{v},\tilde{\bc}^{h}} \big\}$}
{\State 6:~~Update beam set $\bC_{n}= [ \bC_{n-1}, {\bc}^{v}_n \otimes {\bc}^{h}_n ] \in \mathbb{C}^{M \times n}$}
\State 7:~\textbf{end~for}
\State \textbf{Final update}
{\State 8:~Quantized radio paths $\bC=[\bc_1,\cdots,\bc_N] \in \mathbb{C}^{M \times N}$}
{\State 9:~Unquantized weight vector}
{\State ~~~~$\bar{\bz}=\mathfrak{v}_{1} \big\{(\bC^H \bC)^{-1} \bC^H{\bh}{\bh}^H \bC \big\} \in \mathbb{C}^{N}$}
  \end{algorithmic}
\end{algorithm}

{In (\ref{eq_J4:OPBQ}), we do not quantize the weight vector $\bar{\bz}_n \in \mathbb{C}^{n}$ in each update because it is not practical to construct a codebook for weight vectors that constantly change its dimension.} {The problem of choosing the $n$-th {DFT beam}  is then simplified to
\begin{align}
\nonumber
(\bc_{n}^{v},\bc_{n}^{h})&=\argmax_{\tilde{\bc}^{v},\tilde{\bc}^{h} \in  \mathcal{A}_{B_n}^{v} \times \mathcal{A}_{B_n}^{h} } \Bigg( \max_{\tilde{\bz} \in\mathbb{C}^{n}}  \frac{ \big|{\bh}^H \bC_{n|\tilde{\bc}^{v},\tilde{\bc}^{h}} \tilde{\bz}\big|^2}{\big\| \bC_{n|\tilde{\bc}^{v},\tilde{\bc}^{h}} \tilde{\bz}\big\|_2^2} \Bigg)
\\
\nonumber
&\stackrel{(a)}=\argmax_{\tilde{\bc}^{v},\tilde{\bc}^{h} \in  \mathcal{A}_{B_n}^{v} \times \mathcal{A}_{B_n}^{h}  } \mathfrak{eig}_{1} \big\{(\bC_{n|\tilde{\bc}^{v},\tilde{\bc}^{h}}^H \bC_{n|\tilde{\bc}^{v},\tilde{\bc}^{h}})^{-1}
\\
\label{eq_J4:beam}
&~~~~~~~~~~~~~~~~~~~~~~~~~~\bC_{n|\tilde{\bc}^{v},\tilde{\bc}^{h}}^H{\bh}{\bh}^H \bC_{n|\tilde{\bc}^{v},\tilde{\bc}^{h}} \big\} ,
\end{align}
where  $(a)$ is derived from the unquantized weight vector, which is computed based on the generalized Rayleigh quotient \cite{Ref_Bor98}, such that}
\begin{align*}
&\bar{\bz}_{n|\tilde{\bc}^{v},\tilde{\bc}^{h}}=\argmax_{\tilde{\bz} \in \mathbb{C}^n}\frac{  \tilde{\bz}^H (\bC_{n|\tilde{\bc}^{v},\tilde{\bc}^{h}}^H{\bh}{\bh}^H \bC_{n|\tilde{\bc}^{v},\tilde{\bc}^{h}} ) \tilde{\bz} }{\tilde{\bz}^H(\bC_{n|\tilde{\bc}^{v},\tilde{\bc}^{h}}^H\bC_{n|\tilde{\bc}^{v},\tilde{\bc}^{h}}) \tilde{\bz} }
\\
&~~~~~~~=\mathfrak{v}_{1} \big\{(\bC_{n|\tilde{\bc}^{v},\tilde{\bc}^{h}}^H \bC_{n|\tilde{\bc}^{v},\tilde{\bc}^{h}})^{-1} \bC_{n|\tilde{\bc}^{v},\tilde{\bc}^{h}}^H{\bh}{\bh}^H \bC_{n|\tilde{\bc}^{v},\tilde{\bc}^{h}} \big\}.
\end{align*}

{The beam quantization approach gives a set of $N$ quantized beams
\begin{align}
\label{eq_J4:C}
\bC =[\bc_1,\cdots,\bc_N] \in \mathbb{C}^{M \times N}
\end{align}
and the unquantized weight vector
\begin{align}
\label{eq_J4:optz}
\bar{\bz}&= \mathfrak{v}_{1} \big\{(\bC^H \bC)^{-1} \bC^H{\bh}{\bh}^H \bC \big\} \in \mathbb{C}^{N}.
\end{align}
A separate beam quantization should be performed to compute  each codeword candidate based on  Algorithm \ref{Al_J4:01}. In the following Section \ref{sec_J4:Narrowband quantizer}, a practical beam search technique will be proposed  to quantize a single dominant beam with moderate computational complexity as well as compute multiple codeword candidates in a hierarchical fashion.}

{We also evaluate quantization loss due to the beam quantization  as a function of the number of beams  $N$ and the feedback overhead $B_n$ for DFT codebooks. Assuming the unquantized weight vector $\bar{\bz}$, the {beamforming gain} between the channel vector $\bh$ and the set of {DFT beams} $\bC$}
\begin{align}
\nonumber
\mathrm{G}^{\mathrm{bq}} &\doteq \mathrm{E}_{\bh}  { \big[ {|{\bh}^H\bC { \bar{\bz}} |^2}/{\| \bC { \bar{\bz}}\|_2^2} \big]}
\\
\label{eq_J4:bq}
&=\mathrm{E}_{\bh}  { \bigg[ \max_{\tilde{\bz} \in \mathbb{C}^N }\frac{ |{\bh}^H\bC\tilde{\bz} |^2}{\| \bC\tilde{\bz}\|_2^2} \bigg]}
\end{align}
is averaged over channel realizations $\bh$ in {Lemma \ref{lm_J4:bq}.} Before presenting the lemma, we {make} the following assumption.
\begin{assumption}
\label{am:bq}
Assuming a channel vector $\bh$ has already been decomposed into a set of  radio paths $\bD$ and channel gains $\ba$ as in (\ref{eq_J4:simple_channel}), the column vectors for each domain $a \in \{ h,v\}$ in $\bC$ are separately selected as $\bc_{n}^{a} \doteq \ba_{M_a}(q_n/ 2^{B_n}) \in \cA_{B_n}^{a}$, where
\begin{align}
\label{eq_J4:sel_dft}
q_n = \argmax_{q \in \{1,\cdots, 2^{B_n} \}} \big|\bd_{M_a}^H(\psi_{n}^{a})\ba_{M_a}(q/ 2^{B_n})\big|^2.
\end{align}
{Considering   half-wavelength antenna   spacing $d_a=\frac{\lambda}{2}$, the {beamforming gain} between the $n$-th array  response vector and the selected DFT vector  $\Gamma_{na}^2\doteq \mathrm{E}{ \big[ |  \bd_{M_a}^H(\psi_{n}^{a}) \bc_{n}^{a} |^2  \big]}$ is derived  in Appendix \ref{sec_J4:dm}  as}
\begin{align*}
\Gamma_{na}^2 =\frac{1}{M_a^2}\bigg(M_a+ \sum_{q=1}^{M_a-1} \frac{2(M_a-q)\sin{\big(\pi q / 2^{B_n} \big)}}{\pi q / 2^{B_n} }  \bigg).
\end{align*}
\end{assumption}
\begin{lemma}
\label{lm_J4:bq}
{A lower bound of the beamforming gain $\mathrm{G}^{\mathrm{bq}}$  in (\ref{eq_J4:bq}) is approximated as}
\begin{align*}
\mathrm{G}^{\mathrm{bq}} \simeq  \frac{P}{M+N-1}\bigg(N+ \sum_{n=1}^N \sum_{q=n}^P\frac{M \Gamma^2_{nv}\Gamma^2_{nh}-1}{qP} \bigg).
\end{align*}
Please check Appendix \ref{sec_J4:bq} for the proof.
\end{lemma}

\subsection{Phase II: {Beam Combining}}
\label{sec_J4:com_quan}
{In the beam combining phase, we  aim to compute a weight vector $\bz$, which is used to combine beams in $\bC$. To quantize weight vector
\begin{align}
\label{eq_J4:CDA}
\bz&=\argmax_{\tilde{\bz} \in \mathcal{Z}_{B_{c}}} \frac{   |{\bh}^H \bC \tilde{\bz} |^2 }{ \|\bC \tilde{\bz}\|_2^2 },
\end{align}
we  design the codebook including  $U=2^{B_c}$ {unit-norm} combiners $\mathcal{Z}_{B_{c}}\doteq \{ \bz_1,\cdots, \bz_{U} \}$. To study a codebook design framework,  we model the effective channel vector based on the Kronecker correlation model as
\begin{align}
\label{eq_J4:effective_channel}
\bC^H{\bh}   \simeq \bR^{\frac{1}{2}} \bw \in \mathbb{C}^N,
\end{align}
where the covariance matrix $\bR\doteq \mathrm{E}_{\bh} \big[ \bC^H {\bh} {\bh}^H\bC \big] \in \mathbb{C}^{N \times N}$ in (\ref{eq_J4:cov}) is analytically computed in Appendix \ref{sec_J4:cm} and  random variables $\bw =[w_1,\cdots,w_N]^T \in \mathcal{W}$ denotes the weight vector that is subject to the equal gain subset}
\begin{align*}
\mathcal{W} \doteq\big\{ \bw \in \mathbb{C}^N : w_n={e^{j \vartheta_n}}/{\sqrt{N}},~\vartheta_n \sim \mathrm{U}(0,2\pi)\big\}.
\end{align*}

\begin{figure*}[!t]
\normalsize
\centering
\subfigure[$B_{\mathrm{T}}=16$-bits]{\includegraphics[width=0.4955\textwidth]{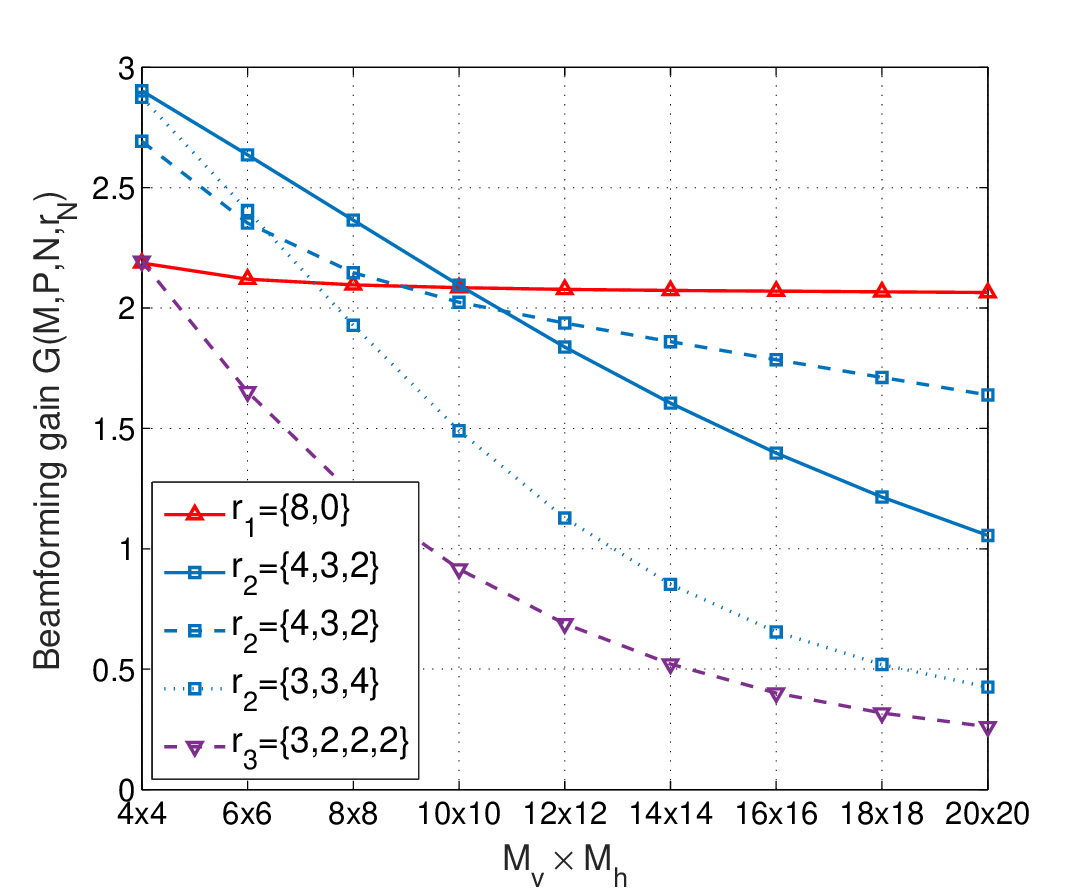}}
\hfil
\subfigure[$B_{\mathrm{T}}=20$-bits]{\includegraphics[width=0.4955\textwidth]{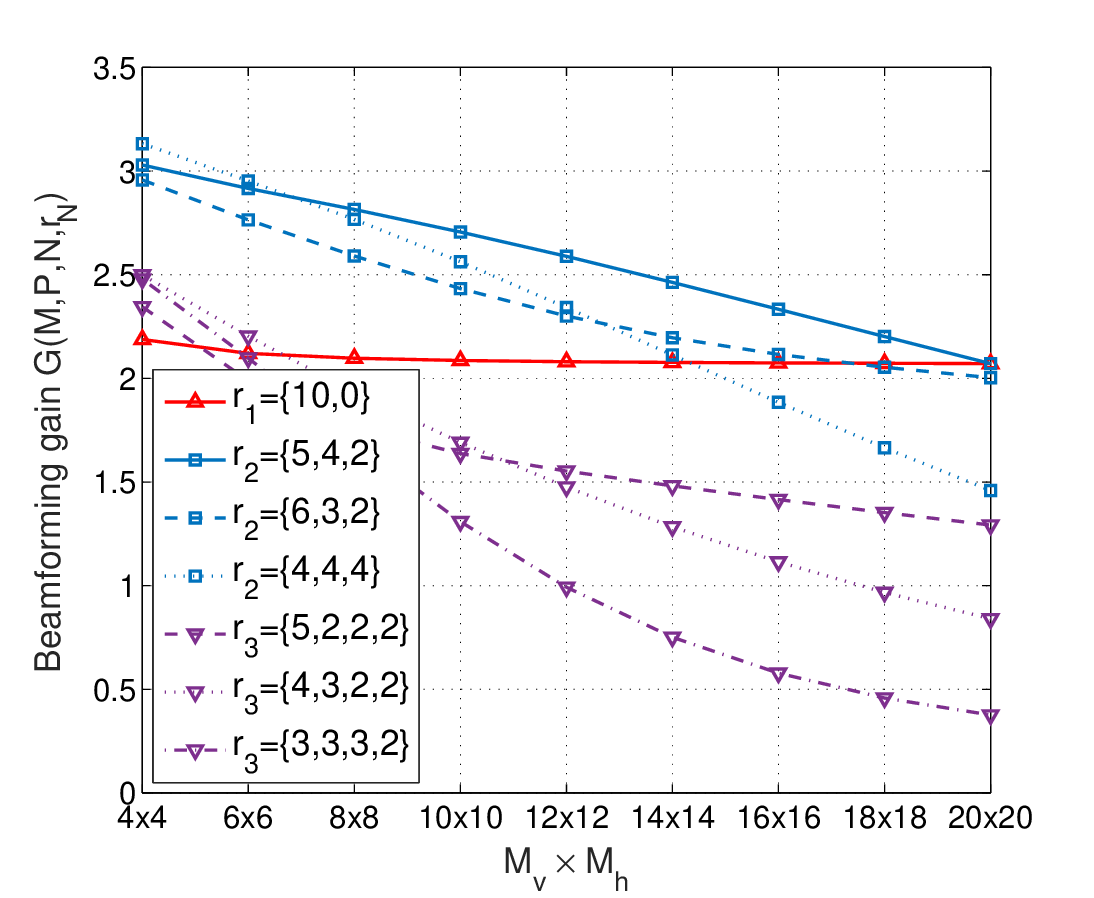}}
\caption{{Cross correlation $\mathrm{G}(M,P,N,\br_N)$ over different feedback-bit allocation  scenarios with $P \in \{3,4,5\}$.}}
\label{fig_J4:anal}
\end{figure*}

{In our codebook design approach, we pick a set of codewords $\{ \bz_1\cdots\bz_{U} \}$  that maximize
\begin{align*}
\sigma_{\mathrm{min}}& = \min_{1\le u \le U} \frac{|\bh^H\bC\bz_u|^2}{\|\bC\bz_u\|_2^2}
\\
&\stackrel{(a)} \geq \min_{1\le u \le U}  \frac{|\bh^H\bC\bz_u|^2}{\| \bC \|_2^2}
\\
&\stackrel{(b)} \simeq \min_{1\le u \le U}  \frac{\big|\bw ^H \big(\bR^{\frac{1}{2}}\big)^H \bz_u \big|^2}{\| \bC \|_2^2},
 \end{align*}
where the inequality in  $(a)$ is  based on  $\|\bC\bz_u \|_2^2 \leq \|\bC \|_2^2 \|\bz_u\|_2^2$ and $\|\bz_u\|_2^2=1$, and $(b)$  is  approximated by plugging the Kronecker correlation model in (\ref{eq_J4:effective_channel}) into the maximizer.}

{Based on the \textit{correlated Grassmannian beamforming} algorithm in \cite{Ref_Lov04}, codewords are then obtained by setting
\begin{align*}
\bz_{u} = {\bR^{\frac{1}{2}}\bee_{u}}/{\| \bR^{\frac{1}{2}}\bee_{u} \|_2}
\end{align*}
and picking a set of  equal gain vectors $\{ \bee_1\cdots\bee_{U} \}$ maximizing
\begin{align*}
{\varrho_{\mathrm{min}}}=\min_{1\le a < b \le U} \sqrt{1-|\bee_a^H\bee_b|^2},
\end{align*}
where equal gain vectors $\bee_u \in \mathcal{E}_N$ are restricted to
\begin{align*}
\mathcal{E}_N \doteq \big\{ \bee \in \mathbb{C}^N : (\bee)_n={e^{j  2\pi \varphi_n}}/{\sqrt{N}},~ \varphi_n \in \big\{{1}/{I} \cdots {I}/{I}\big\} \big\}.
\end{align*}
Note that $I\in \mathbb{N}$ denotes the phase quantization level.}

{We also evaluate the quantization loss due to the beam combining  as a function of the number of codewords $U=2^{B_c}$ in the codebook $\mathcal{Z}_{B_{c}}$ that quantizes the baseband combiner $\bar{\bz}$ in (\ref{eq_J4:optz}).}} To analyze quantization performance of  $\mathcal{Z}_{B_{c}}$, the normalized {beamforming gain} between the normalized effective channel  and the selected {unit-norm} combiner $\bz$
\begin{align}
\nonumber
\mathrm{G}^{\mathrm{bc}} &\doteq \mathrm{E}_{\bh^H\bC}   { \big[ { |{\bh}^H\bC {\bz} |^2}/{ \| \bh^H \bC\|_2^2 } \big]}
\\
\setcounter{equation}{19}
\label{eq_J4:bcp}
& \simeq \mathrm{E}_{\bw}  { \bigg[ \max_{u \in \{1,\cdots,U \}}\frac{ | \bee_u^H\bR\bw |^2}{ (\bw^H\bR\bw) (\bee_u^H \bR \bee_u) } \bigg]}
\end{align}
is averaged over the effective channel  $\bC^H \bh$. Because it is not easy to compute in a closed form, we only derive the normalized {beamforming gain} in the special case of $N=2$ based on the following assumption.
\begin{assumption}
\label{am:bc}
For simple analysis, we assume that the combiner is selected as $\bz \doteq  \bz_{\hat{u}}$, where
\begin{align*}
\hat{u} \doteq \argmax\limits_{u \in \{1,\cdots,U \}} \cos^2 \theta_u.
\end{align*}
Note that $\theta_{u}  \doteq \arccos { |  \bw^H \bee_{u} |}$.
\end{assumption}
\begin{lemma}
\label{lm_J4:bc}
{In the special case of $N=2$, the normalized beamforming gain $\mathrm{G}^{\mathrm{bc}}$  in (\ref{eq_J4:bcp})  is  approximated as}
\begin{align*}
\mathrm{G}^{\mathrm{bc}}  \simeq  \frac{1}{2}\bigg( 1+\frac{U}{\pi} \sin\frac{\pi}{U}\bigg).
\end{align*}
Please check Appendix \ref{sec_J4:bc} for the proof.
\end{lemma}

\subsection{{MIMO Channel Quantizer}}
\label{sec_J4:mu_mimo}
{Before investigating feedback allocation solutions, we extend the proposed quantizer to MIMO channel scenarios. Assuming a MIMO system employing $V$ receive antennas at each user, the channel matrix is defined as
\begin{align*}
\bH_{\textrm{MIMO}}=\big[\bh_1,\cdots,\bh_{V} \big] \in \mathbb{C}^{M \times V}.
\end{align*}
In our MIMO channel quantization approach, DFT beams are chosen to solve the rewritten problem
\begin{align*}
\nonumber
&(\bc_{n}^{v},\bc_{n}^{h})=\argmax_{\tilde{\bc}^{v},\tilde{\bc}^{h} \in  \mathcal{A}_{B_n}^{v} \times \mathcal{A}_{B_n}^{h}  }\mathfrak{eig}_{1} \big\{(\bC_{n|\tilde{\bc}^{v},\tilde{\bc}^{h}}^H \bC_{n|\tilde{\bc}^{v},\tilde{\bc}^{h}})^{-1}
\\
&~~~~~~~~~~~~~~~~~~~~~~~~~~~~~~~~ \bC_{n|\tilde{\bc}^{v},\tilde{\bc}^{h}}^H{\bH}_{\textrm{MIMO}}{\bH}_{\textrm{MIMO}}^H \bC_{n|\tilde{\bc}^{v},\tilde{\bc}^{h}} \big\},
\end{align*}
where the unit-norm combiner is defined by considering possible channel correlations at the receiver according to
\begin{align*}
{\bu}_{n|{\bc}^{v},{\bc}^{h}} = \frac{\bH_{\textrm{MIMO}}^H{\bC_{n|{\bc}^{v},{\bc}^{h}} {\bz}_{n|{\bc}^{v},{\bc}^{h}}}}{ \| \bH_{\textrm{MIMO}}^H{\bC_{n|{\bc}^{v},{\bc}^{h}} {\bz}_{n|{\bc}^{v},{\bc}^{h}}} \|_2} \in \mathbb{C}^V
\end{align*}
and the unquantized weight vector is computed  such as}
\begin{align*}
{\bz}_{n|{\bc}^{v},{\bc}^{h}} &= \mathfrak{v}_{1} \big\{(\bC_{n|{\bc}^{v},{\bc}^{h}}^H \bC_{n|{\bc}^{v},{\bc}^{h}})^{-1}
\\
&~~~~~~~~~~~~~~~ \bC_{n|{\bc}^{v},{\bc}^{h}}^H  {\bH}_{\textrm{MIMO}}{\bH}_{\textrm{MIMO}}^H \bC_{n|{\bc}^{v},{\bc}^{h}} \big\} \in \mathbb{C}^{n}.
\end{align*}

{After constructing the selected set of  DFT beams $\bC$, we  next compute a set of $T$ orthogonal receive combiners and transmit beamformers for spatial multiplexing. For a given set of beams $\bC$, we compute the beamformer for the $t$-th  layer transmission $\bff_t = \bC \bar{\bz}_{t}/ \| \bC \bar{\bz}_{t} \|_2 \in \mathbb{C}^M$, where the unquantized weight vector
\begin{align*}
\bar{\bz}_t=\mathfrak{v}_{t} \big\{(\bC^H \bC)^{-1} \bC^H{\bH_{\textrm{MIMO}}}{\bH^H_{\textrm{MIMO}}}  \bC \big\}  \in \mathbb{C}^{N}
\end{align*}
is computed based on the generalized Rayleigh quotient \cite{Ref_Bor98}. The unit-norm receive combiner is  then given by
\begin{align*}
\bar{\bu}_t=\frac{\bH^H_{\textrm{MIMO}} \bC \bar{\bz}_t}{\| \bH^H_{\textrm{MIMO}} \bC \bar{\bz}_t \|_2} \in \mathbb{C}^{V}.
\end{align*}
The  combining  and precoding matrix are then constructed as
\begin{align*}
\bU &= \big[ \bu_1,\cdots, \bu_T \big] \in \mathbb{C}^{V \times T},
\\
\bF &= \big[ \bff_1,\cdots, \bff_T \big] \in \mathbb{C}^{M \times T},
\end{align*}
respectively, where $T$ denotes the maximum transmission rank.\footnote{Note that designing a beam combining codebook and a feedback resource allocation algorithm that support multi layer MIMO transmission are interesting topics for future research.} Finally, the receive combining  and precoding matrix are multiplied to the left and right side of the channel matrix such as}
\begin{align*}
\bU^H \bH^H_{\textrm{MIMO}} \bF \in \mathbb{C}^{T \times T}.
\end{align*}

\subsection{{Feedback Resource Allocation}}
\label{sec_J4:fba}

In our KP codebook procedure, quantizing more beams {with a high resolution codebook increases the beamforming gain at the cost of} increased feedback overhead. {To effectively allocate limited feedback overhead resources,  {we must} derive the  beamforming gain between the randomly generated channel vectors $\bh$ and the selected codeword $\bC\bz/\|\bC\bz \|_2$ {using}
\begin{align}
\label{eq_J4:cc}
\mathrm{G}  &\doteq \mathrm{E}_{\bh} { \bigg[ \frac{|{\bh}^H\bC\bz |^2}{\| \bC\bz \|_2^2} \bigg]},
\end{align}
{as a function of the feedback-bit allocation scenario\footnote{{In $\br_N$, $B_n$ for $n \in \{1,\cdots,N \}$ denotes the size in bits of the  DFT codebooks $\mathcal{A}_{B_n}^a$ in the domain $a$,  and $B_{c}$ denotes the size in bits of the codebook for combiners $\mathcal{Z}_{B_{c}}$.}} $\br_N $ in (\ref{eq_J4:fba}).} However, inter-dependencies across both quantization phases in Section \ref{sec_J4:beam quantization} and Section \ref{sec_J4:com_quan} make it hard to compute the {beamforming gain} in a closed form. To simplify analysis, we make the following assumption.
\begin{assumption}
\label{am:ge}
{Assuming the  quantization phases in Section \ref{sec_J4:beam quantization} and Section \ref{sec_J4:com_quan} work independently, the channel quantization quality in the proposed KP codebook procedure is evaluated by the combination of the quantization losses in both  phases.}
\end{assumption}

Based on Assumption \ref{am:ge} that both quantization phases are  independent of each other, the beamforming gain in the proposed quantizer is defined by the mixture of $\mathrm{G}^{\mathrm{bq}}$   and $\mathrm{G}^{\mathrm{bc}}$ such as,
\begin{align}
\label{eq_J4:fr}
\mathrm{G}(M,P,N,\br_N)&\doteq\mathrm{G}^{\mathrm{bq}}\mathrm{G}^{\mathrm{bc}}
\\
\nonumber
&= \frac{\mathrm{G}^{\mathrm{bc}}\big(PN+ \sum_{n=1}^N \sum_{q=n}^P\frac{M \Gamma^2_{nv}\Gamma^2_{nh}-1}{q} \big) }{M+N-1},
\end{align}
{where $M$ is the number of  antennas, $P$ is the number of beams in $\bh$, $N$ is the number of dominant beams to be quantized, and $\br_N$ is the feedback-bit allocation scenario in (\ref{eq_J4:fba}).}

{In the proposed   quantizer,  the  feedback scenario are chosen as
\begin{align}
\label{eq_J4:fd}
\big(N,\br_N\big)&=\argmax_{(\tilde{N},\tilde{\br}_{\tilde{N}}) \in \mathcal{N} \times  \mathbb{N}^{\tilde{N}+1}}\mathrm{E}_{P}\big[\mathrm{G}(M,P,\tilde{N},\tilde{\br}_{\tilde{N}}) \big]
\end{align}
by evaluating all possible scenarios $\tilde{\br}_{\tilde{N}} = [\tilde{B}_1,\cdots,\tilde{B}_{\tilde{N}},\tilde{B}_c ]^T$ that considers $\tilde{N}$ beams in  $\mathcal{N} \doteq \{ 1,2,3 \}$. Note that the possible feedback  scenarios are subject to the total feedback overhead $B_{\mathrm{T}}=\tilde{B}_c+\sum_{n=1}^{\tilde{N}} 2\tilde{B}_{n}$-bits.} In (\ref{eq_J4:fd}), the expectation is taken over the number of dominant paths $P$ since $P$ varies depending on the channel environments. By assuming  $P$ is equally probable from $3$ to $5$, we plot the arithmetic mean of $\mathrm{G}(M,P,\tilde{N},\br_{\tilde{N}}) $ in Fig. \ref{fig_J4:anal} with different numbers of antennas and feedback bits.

As shown in the figure, quantizing one or two beams give the best performance under practical UPA scenarios and feedback overheads. {Therefore, we construct the codebook $\cF_1$ for quantizing a single 2D DFT beam and the codebook  $\cF_2$ combining two quantized 2D DFT beams  based on the predefined feedback-bit allocation scenarios\footnote{{The total feedback overhead for feedback scenarios are $B_{\mathrm{T}}  = 2B_1 + 2 \check{B}_1$-bits and $B_{T}= 2B_1 +2B_2+B_c$-bits, where $2\check{B}_1=2B_2+B_c$.}}
\begin{align}
\label{eq_J4:FRA}
\br_{1}= [B_1+\check{B}_1,0]\in \mathbb{N}^2 ,~~~\br_2 =[B_1,B_2,B_c] \in \mathbb{N}^3,
\end{align}
respectively. The final codebook is then defined such that}
\begin{align*}
\cF \doteq \cF_1 \cup \cF_2.
\end{align*}
\begin{remark}
\label{rm_J4:02}
{In most of channel realizations, the inter-user interference  due to the remained paths is negligible because most of channel gains are contained in the first and second dominant beams \cite{Ref_Son16_con,Ref_Cho15}. Based on the codebook subset restriction algorithm in \cite{Ref_Rum13,Ref_Cha15}, severe inter-user interference could be mitigated by   reporting the remained paths having a considerable amount of channel gains.}
\end{remark}

\subsection{{Beam Search Approach}}

\label{sec_J4:Narrowband quantizer}

{It is necessary to search both vertical and horizontal domains jointly to scan {for the} dominant beams in a channel vector.} However, this joint approach increases a computational complexity. For example, it is required to carry out $2^{2(B_1+\check{B}_1)}$ vector computations to scan a single {2D DFT beam} under the  feedback-bit allocation scenario $\br_{1}$ in (\ref{eq_J4:FRA}). To reduce the heavy computational complexity that comes with detecting the single dominant beam, we propose a multi-round beam search technique as follows.

\textbf{Round 1}: {For} the channel vector $\bh$, the first dominant beam is chosen  using  DFT codebooks  $\mathcal{A}_{B_1}^{v}$ and $\mathcal{A}_{B_1}^{h}$ in (\ref{eq_J4:DFTCB}), which have low-resolution DFT vectors. The DFT beam is  given by
\begin{align}
\label{eq_J4:first}
{\bc}_{1} &= {\bc}_{1}^{v} \otimes {\bc}_{1}^{h},
\\
\nonumber
({\bc}_{1}^{v},{\bc}_{1}^{h}) &= \argmax_{ (\tilde{\bc}^{v},\tilde{\bc}^{h}) \in \mathcal{A}_{B_1}^{v} \times \mathcal{A}_{B_1}^{h}}|{\bh}^H(\tilde{\bc}^{v}  \otimes \tilde{\bc}^{h}   )|^2.
\end{align}
{Later, the selected 2D DFT beam in (\ref{eq_J4:first}) will be a baseline that guides the generation of two codeword candidates.}

\textbf{Round 2}:  
{In this round, $2\check{B}_1$-bits  are assigned for constructing  the two codeword candidates.

1) To support channel realizations having a single dominant beam, a codeword is computed {based on the feedback-bit allocation scenario $\br_{1}$ in (\ref{eq_J4:FRA}) by scanning beam directions near  ${\bc}_{1}$ from  Round 1.}  The first codeword is given by
\begin{align*}
{\bff}_{1}   &= \check{\bd}_{M}(  {{\theta}_{1}^{v}}, {{\theta}_{1}^{h}} ) \odot \bc_1,
\\
({\theta}_{1}^{v},{\theta}_{1}^{h}) &= \argmax_{ \tilde{\theta}^{v},\tilde{\theta}^{h}  \in \mathcal{T}_{B_1}^{\check{B}_1} } \big|{\bh}^H  \big(   \check{\bd}_{M}(\tilde{\theta}^{v} , \tilde{\theta}^{h})   \odot \bc_1\big)  \big|^2
\end{align*}
where  $\check{\bd}_{M}(\theta^{v}, \theta^{h} )\doteq \sqrt{M} \bd_{M}(\theta^{v} , \theta^{h} )$ is defined to shift {the} beam  directions\footnote{The Hadamard product formulation satisfies the following formulation $\sqrt{M}\bd_{M}(\theta_{1},\theta_{3}) \odot \bd_{M}(\theta_{2},\theta_{4})=\bd_{M}(\theta_{1}+\theta_{2},\theta_{3}+\theta_{4}).$} and the $\check{B}_1$-bit size codebook
\begin{align}
\label{eq_J4:ZZ}
\mathcal{T}_{B_1}^{\check{B}_1}=\bigg\{-\frac{1-2^{-\check{B}_1}}{ 2^{B_1+1}} :  {2^{-(B_1+\check{B}_1)}} : \frac{1-2^{-\check{B}_1}}{ 2^{B_1+1}} \bigg\}
\end{align}
is designed for refining beam directions of any {DFT beams}.

{In our CSI quantization approach, the first codebook is then defined such that
\begin{align*}
\cF_1 \doteq \big\{  {\bff}_1 \in \mathbb{C}^M :  {\bff}_1= (\tilde{\bc}_{1}^{v} \otimes \tilde{\bc}_{1}^{h}) \odot \check{\bd}_{M}(\tilde{\theta}^{v} , \tilde{\theta}^{h})  \big\}
\end{align*}
over $\tilde{\bc}_{1}^{v} \in \mathcal{A}_{B_1}^{v}$, $\tilde{\bc}_{1}^{h} \in \mathcal{A}_{B_1}^{h}$, and $\tilde{\theta}^{v},\tilde{\theta}^{h} \in \mathcal{T}_{B_1}^{\check{B}_1}$.}

2) To support channel realizations having multiple dominant beams, a codeword is computed {based on the feedback-bit allocation scenario $\br_{2}$ in (\ref{eq_J4:FRA})} by choosing  an additional DFT beam {to combine} with ${\bc}_{1}$.  The second codeword is given by
\begin{align*}
\bff_{2} &= \frac{ [{\bc_1,{\bc}_{2}^{v} \otimes {\bc}_{2}^{h}}] {\bz} }{  \| [{\bc_1,{\bc}_{2}^{v} \otimes {\bc}_{2}^{h}}] {\bz}\|_2},
\\
( {\bc}_{2}^{v},{\bc}_{2}^{h},{\bz}) & =\argmax_{ (\tilde{\bc}^{v},\tilde{\bc}^{h},\tilde{\bz}) \in \mathcal{A}_{B_2}^{v} \times  \mathcal{A}_{B_2}^{h} \times \mathcal{Z}_{{B_{c}}} } { \bigg|  \frac{ \bh^H[{\bc_1,\tilde{\bc}^{v} \otimes \tilde{\bc}^{h}} ]\tilde{\bz}}{\|  [{\bc_1,\tilde{\bc}^{v} \otimes \tilde{\bc}^{h}} ] \tilde{\bz}  \|_2} \bigg|^2}
\end{align*}
using} $B_2$-bits size DFT codebooks  $\mathcal{A}_{B_2}^{v}$ and $\mathcal{A}_{B_2}^{h}$ and $B_{c}$-bits size codebook $\mathcal{Z}_{B_{c}}$, {which is developed to combine the {two {2D DFT beams} as explained in the Section \ref{sec_J4:com_quan}.}

{Considering our CSI quantization technique, the second codebook is then defined such that
\begin{align*}
\cF_2 \doteq \bigg\{  {\bff}_2 \in \mathbb{C}^M :  {\bff}_2= \frac{\big[\tilde{\bc}_{1}^{v} \otimes \tilde{\bc}_{1}^{h},\tilde{\bc}_{2}^{v} \otimes \tilde{\bc}_{2}^{h}\big]\tilde{\bz}}{\big\| \big[\tilde{\bc}_{1}^{v} \otimes \tilde{\bc}_{1}^{h},\tilde{\bc}_{2}^{v} \otimes \tilde{\bc}_{2}^{h}\big] \tilde{\bz}\big\|_2}   \bigg\}
\end{align*}
over $\tilde{\bc}_{1}^{v} \in \mathcal{A}_{B_1}^{v}$, $\tilde{\bc}_{1}^{h} \in \mathcal{A}_{B_1}^{h}$, $\tilde{\bc}_{2}^{v} \in \mathcal{A}_{B_2}^{v}$, $\tilde{\bc}_{2}^{h} \in \mathcal{A}_{B_2}^{h}$, and $\tilde{\bz} \in \mathcal{Z}_{{B_{c}}} $.}

{\textbf{Round 3}}: {Using the} two codeword candidates $\bff_1 \in \cF_1$ and $\bff_2 \in \cF_2$, the final codeword is selected with an additional bit
\begin{align}
\label{eq_J4:final code}
\bff=\argmax_{\tilde{\bff} \in \{\bff_1,\bff_2\}} \big| \bh^H\tilde{\bff} \big|^2.
\end{align}

\section{Proposed Wideband Quantizer}

\begin{figure}[!t]
\centering
\subfigure[Wideband resource blocks]{\label{fig_J4:WRB}\includegraphics[width=0.4955\textwidth]{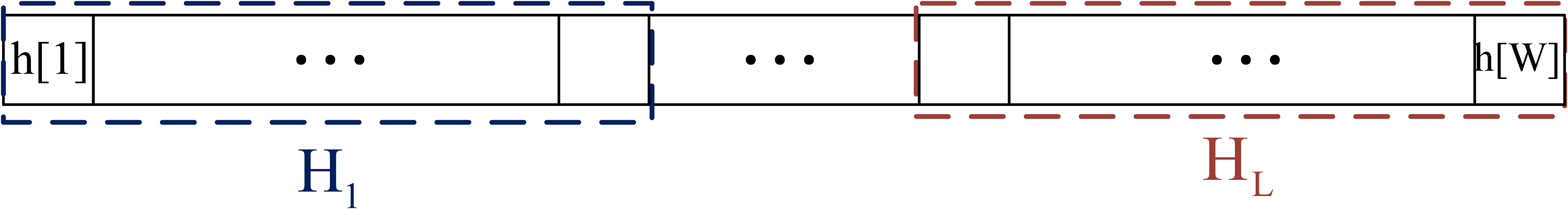}}
\hfil
\subfigure[Narrowband resource blocks]{\label{fig_J4:NRB}\includegraphics[width=0.4955\textwidth]{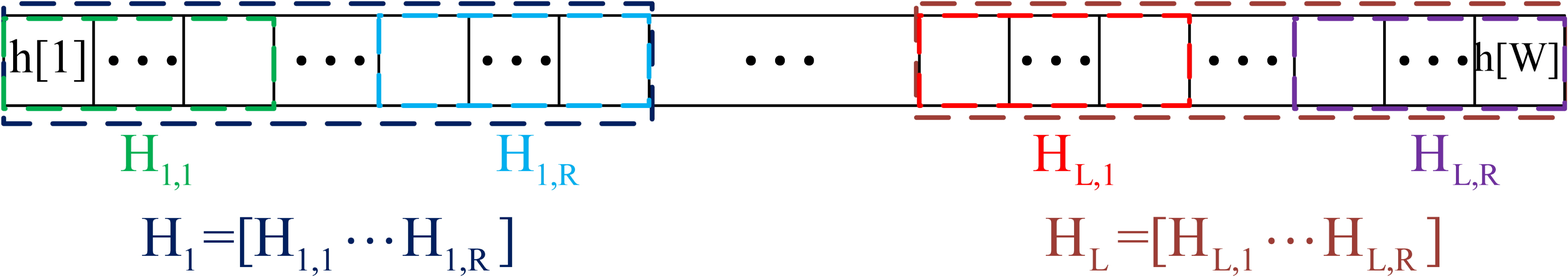}}
\caption{An overview of wideband model having multiple tones.}
\label{fig_J4:RB}
\end{figure}

We develop a wideband quantizer that takes multiple frequency tones into account. 
Before developing practical quantizers, we overview a broadband {system model adopted in 3GPP LTE-Advanced.} As shown in Fig. \ref{fig_J4:WRB}, $W$ total frequency tones are divided into $L$ wideband {RBs where each wideband RB includes} $\frac{W}{L}$ channels. Each wideband RB  is  written {in  a matrix form as}
\begin{align*}
\bH_{\ell}= \big[ \bh [1+{(\ell-1)W}/{L} ],\cdots, \bh[ {\ell}W/{L}] \big] \in \mathbb{C}^{M \times \frac{W}{L}}.
\end{align*}
{As depicted in Fig. \ref{fig_J4:NRB}, each wideband RB is divided into $R$ narrowband RBs
\begin{align*}
\bH_{\ell} =\big[ \bH_{\ell,1},\cdots,\bH_{\ell,R} \big],
\end{align*}
where $\bH_{\ell,r} \in \mathbb{C}^{M \times \frac{W}{LR}}$ denotes the narrowband RB that is written in a matrix form.}

Next,  {the correlation between the channel vectors} is studied numerically based on the cross correlations over {$w_1 \ne w_2$}
\begin{align*}
\gamma_{\bh} &\doteq \mathrm{E}_{\bh[w]} \bigg[ \frac{\big| {\bh[w_1]^H\bh[w_2]}\big|^2}{\| \bh[w_1] \|_2^2 \| \bh[w_2] \|_2^2}\bigg],
\\
\gamma_{\bc_1}&\doteq \mathrm{E}_{\bh[w]}  \big[ \big| \bc_1[w_1]^H\bc_1[w_2]\big|^2 \big],
\end{align*}
where $\bh[w]$ denotes 3D-SCM channel vectors and $\bc_1[w]$ denotes the dominant {2D DFT beam} that is chosen from the DFT codebooks $\mathcal{A}_{B}^{v}$ and $\mathcal{A}_{B}^{h}$ in (\ref{eq_J4:DFTCB}) {for subcarrier $w$.} As shown  in Fig. \ref{fig_J4:compa}, it is verified that the dominant {2D DFT beams} in the different subcarriers' channel vectors are highly correlated. {Based on empirical studies, the wideband quantizer is designed in such a way that the correlated information, i.e., the dominant {2D DFT beam}, is shared between neighboring subcarriers.} 

{\textbf{Level 1 (Wideband resource block)}}: We choose two {2D DFT beams} {that are close to the} channel vectors in each wideband RB. For supporting the $\ell$-th wideband RB, the first {2D DFT beam} is chosen as
\begin{align}
{\bc}_{1|\ell} &= {\bc}_{1}^{v} \otimes {\bc}_{1}^{h},
\\
\nonumber
({\bc}_{1}^{v},{\bc}_{1}^{h})&= \argmax_{(\tilde{\bc}^{v},\tilde{\bc}^{h}) \in {\mathcal{A}_{B_{\textrm{W1}}}^{v}} \times  {\mathcal{A}_{B_{\textrm{W1}}}^{h}}} \big\| \bH_{\ell}^H(\tilde{\bc}^{v}  \otimes \tilde{\bc}^{h}  )\big\|_2^2
\end{align}
{with $B_{\textrm{W1}}$-bit DFT codebooks. Next, the second  {2D DFT beam} is chosen using $B_{\textrm{W2}}$-bit DFT codebooks as}
\begin{align}
{\bc}_{2|\ell} &={\bc}_{2}^{v} \otimes {\bc}_{2}^{h},
\\
\nonumber
({\bc}_{2}^{v},{\bc}_{2}^{h})&=\argmax_{\tilde{\bc}^{v},\tilde{\bc}^{h}} \max_{\tilde{\bz} \in \mathcal{Z}_{B_c}} \bigg\|  \frac{{\bH}_\ell^H [{\bc}_{1|\ell},\tilde{\bc}^{v} \otimes \tilde{\bc}^{h} ]\tilde{\bz}}{\| [{{\bc}_{1|\ell},\tilde{\bc}^{v} \otimes \tilde{\bc}^{h}}]\tilde{\bz}  \|_2}  \bigg\|_2^2
\end{align}
{over $(\tilde{\bc}^{v},\tilde{\bc}^{h}) \in \mathcal{A}_{B_{\textrm{W2}}}^{v} \times \mathcal{A}_{B_{\textrm{W2}}}^{h}$.}

\begin{figure}[!t]
\centering
\subfigure{\includegraphics[width=0.4355\textwidth]{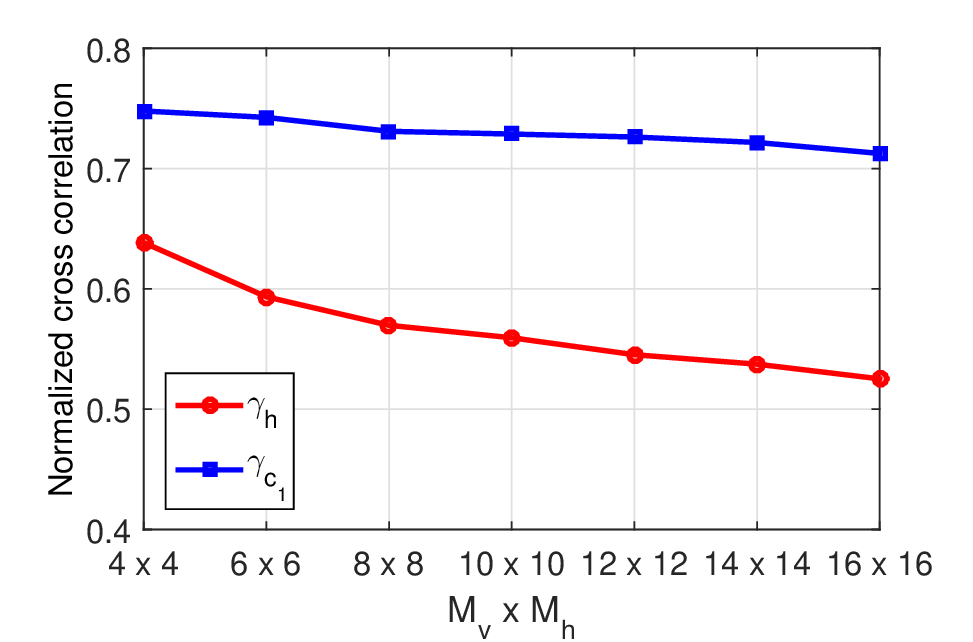}}
\caption{{Normalized {beamforming gain}s between subcarrier channel  vectors. $(B=5,~d_v=0.8 \lambda,~d_h=0.5 \lambda,~ w \in \{1,\cdots,150\})$}}
\label{fig_J4:compa}
\end{figure}

\begin{figure*}[!t]
\normalsize
\centering
\subfigure[$B_{1}=4,~B_{2}=4,~B_{c}=2~(B_{\textrm{T}}=18)$]{\includegraphics[width=0.4655\textwidth]{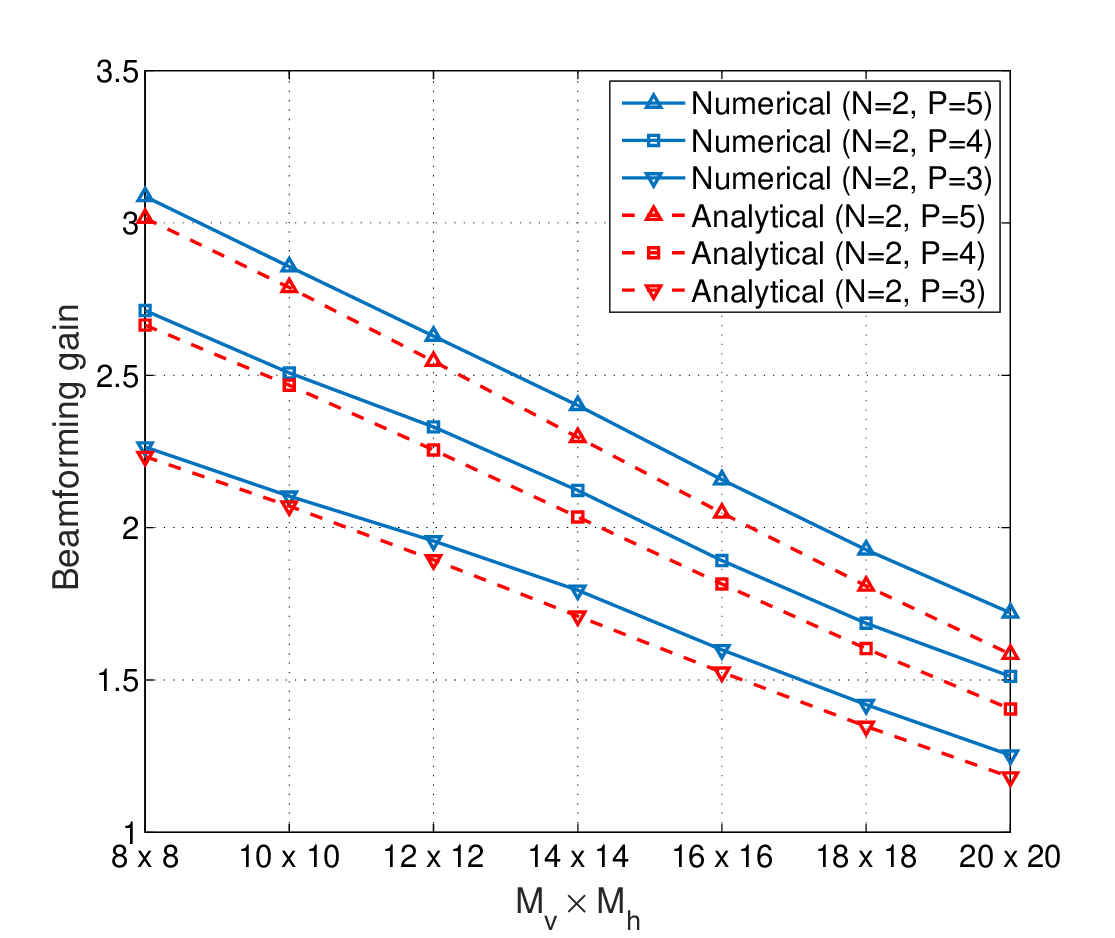}}
\hfil
\subfigure[$B_{1}=5,~B_{2}=4,~B_{c}=2~(B_{\textrm{T}}=20)$]{\includegraphics[width=0.4655\textwidth]{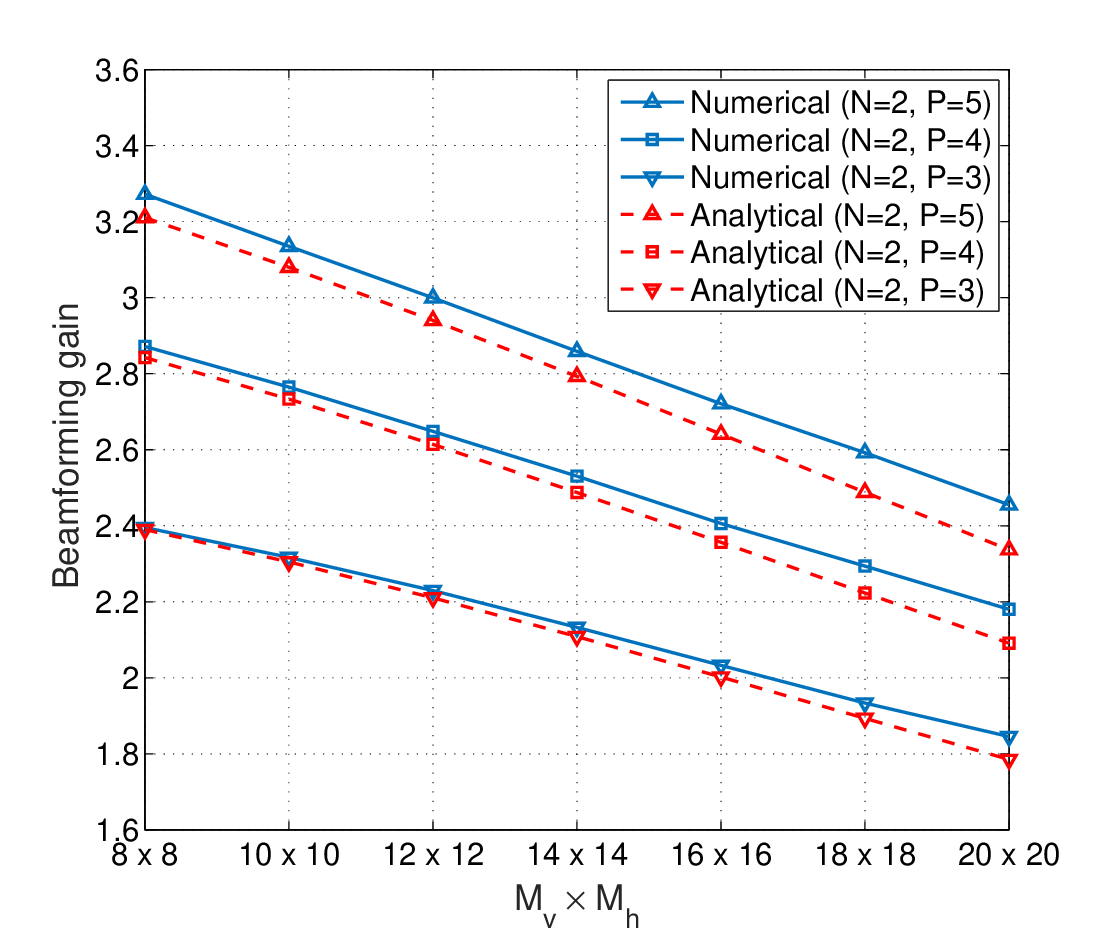}}
\caption{{Beamforming gain comparison between the numerical results $\mathrm{G}$ in (\ref{eq_J4:cc}) and the analytical results $\mathrm{G}(M,P,N,\br_N)$ in (\ref{eq_J4:fr}).}}
\label{fig_J4:NBG}
\end{figure*}

\begin{figure*}[!t]
\normalsize
\centering
\subfigure[$d_v=0.8 \lambda,~d_h=0.5 \lambda$]{\label{fig_J4:single_a}\includegraphics[width=0.4955\textwidth]{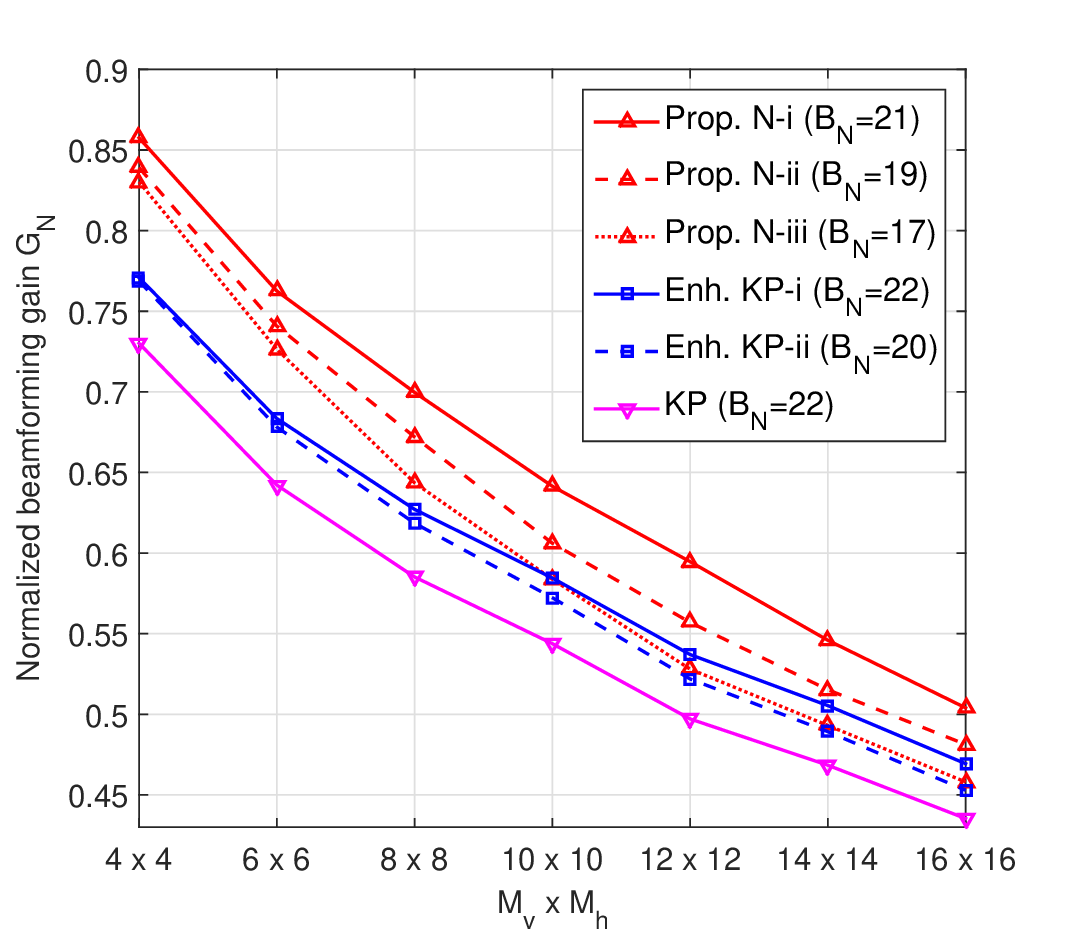}}
\hfil
\subfigure[$d_v=1.6 \lambda,~d_h=1.0 \lambda$]{\label{fig_J4:single_b}\includegraphics[width=0.4955\textwidth]{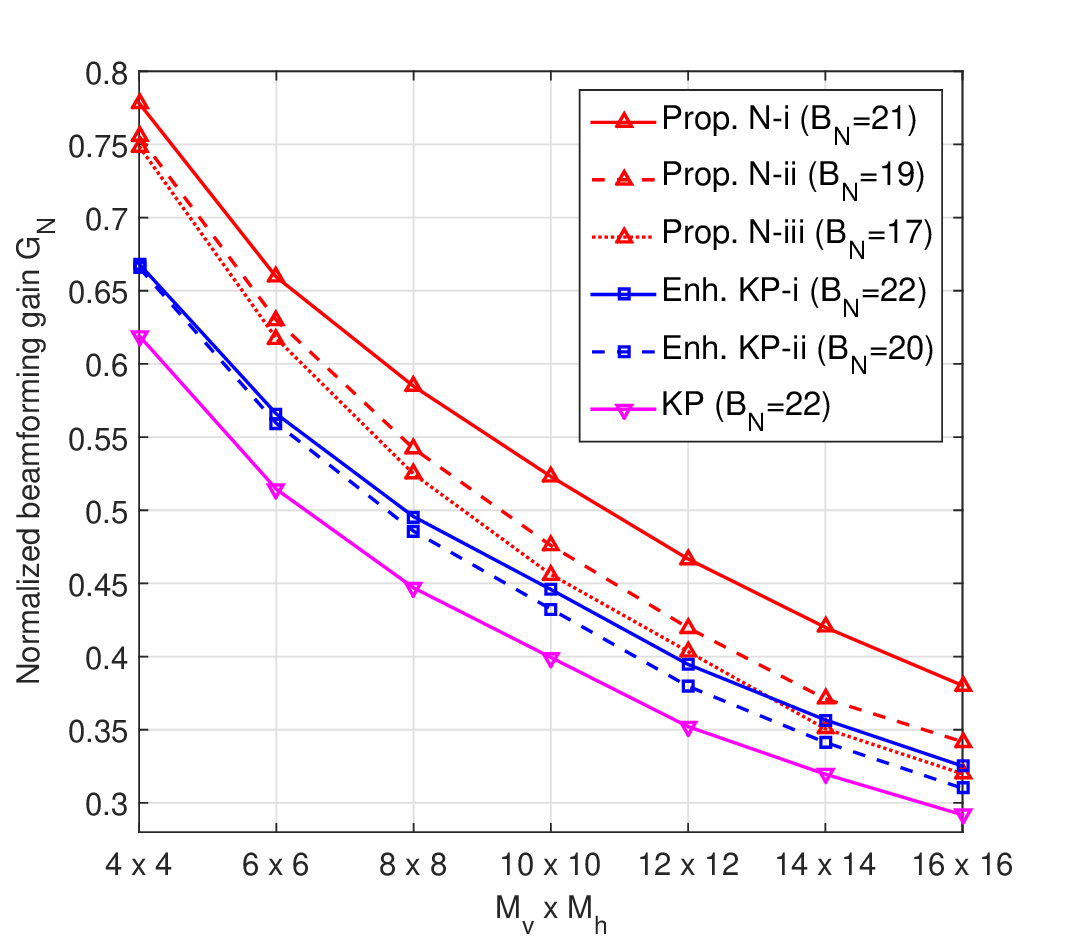}}
\caption{Normalized beamforming gain of narrowband quantizers.}
\label{fig_J4:single}
\end{figure*}

{\textbf{Level 2 (Narrowband resource block)}}: Within the $\ell$-th wideband RB, the set of {2D DFT beams} $\bc_{1|\ell}$ and $\bc_{2|\ell}$  will be a baseline that guide the quantization of channel vectors in each narrowband RB. To construct each set of two codeword candidates, $2B_{\textrm{N1}}$-bits are allocated for each narrowband RB.

\textbf{Round 1}: The first codeword is computed to support the channel scenario having a single dominant beam. The first codeword quantizes {channel vectors $\bH_{\ell,r}$} in the $r=\lceil R(\frac{wL}{W}-\ell-1) \rceil$-th narrowband RB of the $\ell= \lceil \frac{wL}{W} \rceil$-th wideband RB  by refining the beam direction of $\bc_{1|\ell}$ {according to}
\begin{align*}
{\bff}_{1|\ell,r} &= \check{\bd}_{M}({\theta}_{1}^{v} ,{\theta}_{1}^{h})\odot \bc_{1|\ell},
\\
({\theta}_{1}^{v},{\theta}_{1}^{h}) &= \argmax_{\tilde{\theta}^{v},\tilde{\theta}^{h} \in \mathcal{T}^{B_{\textrm{N1}}}_{B_{\textrm{W1}}} } \big\| \bH^H_{\ell,r}    \big(   \check{\bd}_{M}(\tilde{\theta}^{v},\tilde{\theta}^{h}) \odot \bc_{1|\ell} \big)  \big\|_2^2
\end{align*}
{with $B_{\textrm{N1}}$-bit codebooks in (\ref{eq_J4:ZZ}).}

\textbf{Round 2}: The second codeword is computed to support the channel scenario  having two dominant beams. The proposed quantizer only refines the direction of ${\bc}_{1|\ell}$ as well as combines with the second {2D DFT beam} ${\bc}_{2|\ell}$. The second codeword is
\begin{align*}
{\bff}_{2|\ell,r} &= \frac{  \big[ \check{\bd}_{M}({\theta}_{2}^{v},{\theta}_{2}^{h}) \odot \bc_{1|\ell},\bc_{2|\ell} \big] {\bz} }{\big\| \big[ \check{\bd}_{M}({\theta}_{2}^{v},{\theta}_{2}^{h}) \odot \bc_{1|\ell},\bc_{2|\ell} \big] {\bz}  \big\|_2},
\\
({\theta}_{2}^{v},{\theta}_{2}^{h},{\bz_2}) &=\argmax_{\tilde{\theta}^{v},\tilde{\theta}^{h},\tilde{\bz} }{\Bigg\|  \frac{ \bH^H_{\ell,r}\big[ \check{\bd}_{M}(\tilde{\theta}^{v},\tilde{\theta}^{h}) \odot \bc_{1|\ell},\bc_{2|\ell} \big]  \tilde{\bz}}{\big\| \big[ \check{\bd}_{M}(\tilde{\theta}^{v},\tilde{\theta}^{h}) \odot \bc_{1|\ell},\bc_{2|\ell} \big]  \tilde{\bz} \big\|_2} \Bigg\|_2^2}
\end{align*}
over $\tilde{\theta}^{v},\tilde{\theta}^{h} \in \mathcal{T}^{B_\textrm{N2}}_{B_\textrm{W1}}$  in (\ref{eq_J4:ZZ}) and $\tilde{\bz} \in \mathcal{Z}_{B_{c}}$. Out of $2B_{\textrm{N1}}$-bits allocated for each narrowband RB, $2B_{\textrm{N2}}$ bits are assigned for refining the first {2D DFT beam} and $B_{c}$ bits are assigned for combining the two {2D DFT beams}.

\begin{table}
\caption{3D-SCM simulation parameters.}
\centering
\begin{tabular}{|c|c|}
  \hline
 Tx antennas& $4\times4$ to $20\times20$ Co-polarized UPA\\ \hline
  Rx antennas& $1$ Co-polarized UPA\\ \hline
 Scenario & UMi NLOS   \\ \hline
 Carrier frequency $f_c$ & $2$ GHz   \\ \hline
 Subcarrier spacing $\triangle$  & $15$ KHz   \\ \hline
 Vertical antenna spacing $d_v$ & $0.8\lambda_c,~1.6\lambda_c$    \\ \hline
  Horizontal antenna spacing $d_h$ & $0.5\lambda_c,~1.0\lambda_c$  \\ \hline
\end{tabular}
\label{tab:sp}
\end{table}

\textbf{Round 3}: Among the two codeword candidates, the final codeword is selected with an additional bit {according to}
\begin{align}
\label{eq_J4:sel_dual}
\bff_{\ell,r} = \argmax_{\tilde{\bff} \in \{ \bff_{1|\ell,r},\bff_{2|\ell,r} \}} \big\| \bH^H_{\ell,r} \tilde{\bff} \big\|^2_2.
\end{align}

\section{Simulation Results}
\label{sec_J4:Sim}

We verify the performance of our CSI quantizers. Before evaluating the proposed quantizers, we pause to validate the  accuracy of the approximated {beamforming gain} in (\ref{eq_J4:fr}). The {beamforming gain} between the simplified channel ${\bh}$ and the quantized  channel ${\bC\bz}/{\|\bC\bz \|_2}$ is computed as in (\ref{eq_J4:cc}). 
For numerical simulations, $10,000$ channels  in (\ref{eq_J4:simple_channel}) are generated by assuming a fixed number of the ray-like beams $P \in \{3,4,5 \}$. {In Fig. \ref{fig_J4:NBG}, it is shown that the approximated formula in (\ref{eq_J4:fr}) gives a tight lower bound on the numerical results in (\ref{eq_J4:cc}).}

\begin{figure*}[!t]
\normalsize
\centering
\subfigure[$d_v=0.8 \lambda,~d_h=0.5 \lambda$]{\label{fig_J4:wide_a}\includegraphics[width=0.4955\textwidth]{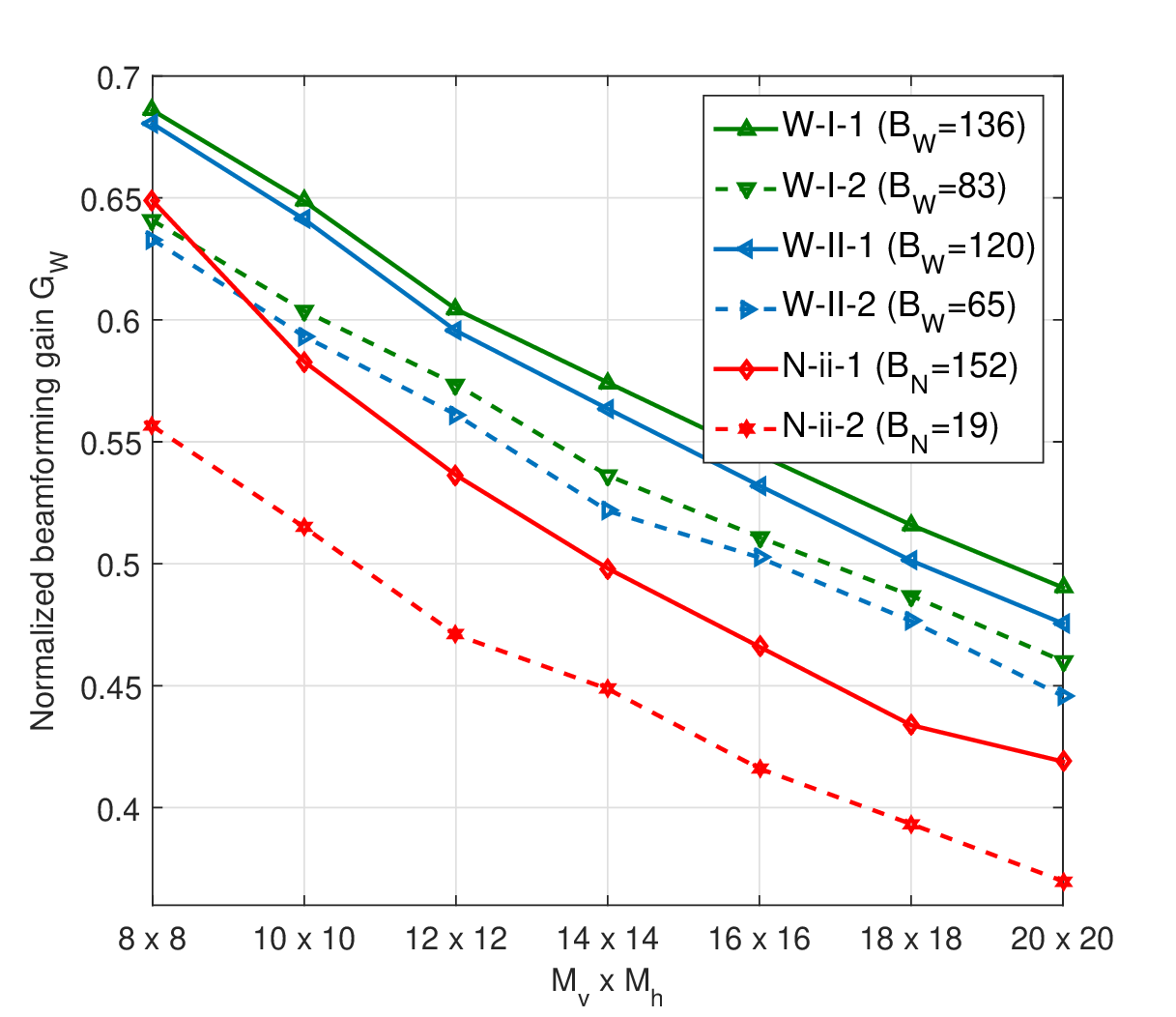}}
\hfil
\subfigure[$d_v=1.6 \lambda,~d_h=1.0 \lambda$]{\label{fig_J4:wide_b}\includegraphics[width=0.4955\textwidth]{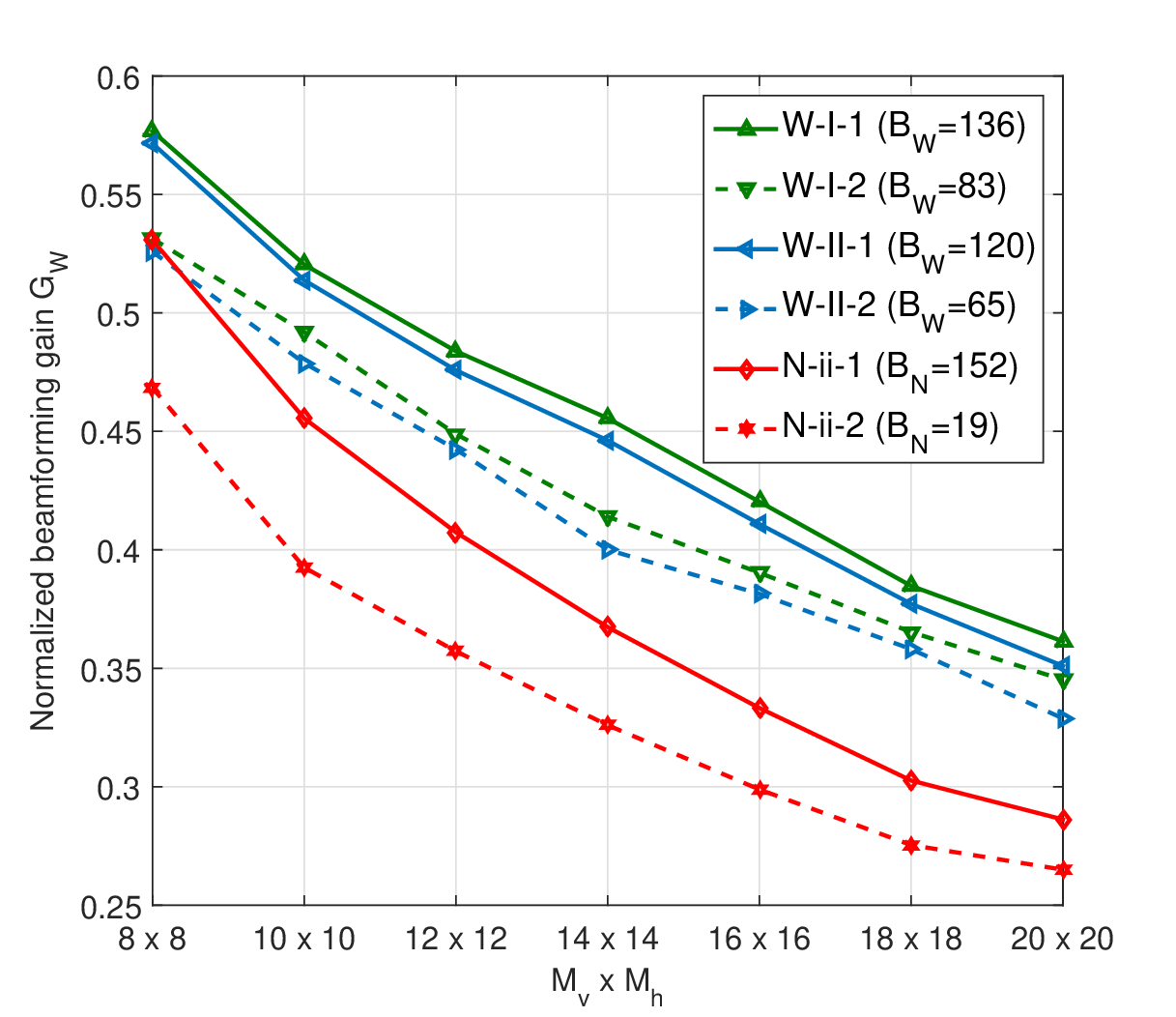}}
\caption{{Normalized beamforming gain comparison between the narrowband and wideband quantizers.}}
\label{fig_J4:wide}
\end{figure*}

We now evaluate the narrowband quantizers {using simulations.} Numerical results are obtained through Monte Carlo simulations with $10,000$ channel realizations. For generating 3D-SCM channels, we use the  parameters in Table \ref{tab:sp}. We evaluate the normalized beamforming gain of narrowband quantizer
\begin{align*}
\mathrm{G}_{\textrm{N}}\doteq \mathrm{E}_{\bh} \bigg[  \frac{|\bff^H {\bh}|^2}{\|\bh \|_2^2}   \bigg],
\end{align*}
where $\bff$ is the final codeword\footnote{{The perfect CSI beamformer gives the normalized beamforming gain of one.}} that is chosen in (\ref{eq_J4:final code}). We {compare the beamforming} gains of the proposed quantizer with that of  the KP codebooks in \cite{Ref_3GPP150057,Ref_3GPP150560} and the enhanced KP codebook in \cite{Ref_Cho15}. The {feedback-bit allocation}\footnote{{The {feedback-bit allocation} scenarios for the proposed narrowband quantizers  are predefined in (\ref{eq_J4:FRA}).}} of each quantizer is listed in Table \ref{tab:sq} and the computational complexity\footnote{We count the number of vector computations to evaluate the complexity.} $V_{\textrm{N}}$ and  feedback overhead\footnote{The feedback overhead (per each frequency tone CSI) is assessed by the combination of the overheads for both the first and second rounds in Section \ref{sec_J4:Narrowband quantizer}.} $B_{\textrm{N}}$ are summarized in Table \ref{tab:complex}.

\begin{table}
\caption{Feedback configurations of narrowband quantizer}
\centering
\begin{tabular}{|c|l|l|c|c|}
  \hline
  &  $1^{\textrm{st}}$ round  &    $2^{\textrm{nd}}$ round & $B_{\textrm{N}}$ & $V_{\textrm{N}}$ \\ \hline
    Prop. N-i  &     $B_{1}=5$    &$\check{B}_{1}=5,~B_{2}=4,B_{c}=2 $  &  $21$ & 3,072 \\ \hline
    Prop. N-ii  &     $B_{1}=5$    &$\check{B}_{1}=4,~B_{2}=3,B_{c}=2 $&  $19$ & 1,536 \\ \hline
 Prop. N-iii  &     $B_{1}=4$    &$\check{B}_{1}=4,~B_{2}=3,B_{c}=2$  &  $17$ & 768 \\ \hline
   Enh. KP-i  &     $B_{1}+B_2$    &$B_{1}=5,~B_{2}=5$  &  $22$ & 2,176  \\ \hline
  Enh. KP-ii  &     $B_{1}+B_2$    &$B_{1}=5,~B_{2}=4$  &  $20$ & 1,120  \\ \hline
KP codebook  &     $B_{1}=11$    &  &  $22$  & 4,096 \\ \hline
\end{tabular}
\label{tab:sq}
\end{table}

\begin{table}
\caption{Feedback overheads and complexity comparisons}
\centering
\begin{tabular}{|c|c|c| }
  \hline
     & Feedback overhead $B_{\textrm{N}}$ & Vector computations $V_{\textrm{N}}$   \\ \hline
       Prop. quantizer  &  $2(B_{1}+B_{2})+B_c+1$ & $2^{2B_1}+2^{2B_2+B_c+1}$   \\ \hline
      Enhanced KP &  $2(B_1+B_2+1)$ & $2\big(2^{B_{1}+B_{2}}+2^{B_{1}}+2^{B_{2}} \big)$\\  \hline
       KP codebook&  $2B_1$ & $2^{B_1+1}$\\  \hline
\end{tabular}
\label{tab:complex}
\end{table}

In Figs. \ref{fig_J4:single_a} and \ref{fig_J4:single_b}, the normalized beamforming gains of {the three quantizers are} plotted with different antenna spacing scenarios.  The proposed quantizer searches both vertical and horizontal domains jointly, while other KP codebooks search beams lying in each domain independently and integrate the results later. The {2D DFT beams}, which are quantized in the proposed quantizer, are aligned by {cophasing and scaling each beam.} {On the contrary, the} quantized beams in the enhanced KP codebook  \cite{Ref_Cho15}  are simply added up together without considering phase alignment. For these reasons, the proposed quantizer generates  higher beamforming gains than {those of other KP codebooks.}

\begin{table}
\caption{{Feedback overheads for each wideband and narrowband RB}}
\centering
\begin{tabular}{|c|c|c|}
  \hline
  &Level $1$: Wideband RB & Level $2$: Narrowband RB  \\ \hline
     W-I  &   $B_{\textrm{W1}}=5,~B_{\textrm{W2}}=5$    & $B_{\textrm{N1}}=3,B_{\textrm{N2}}=2,B_{c}=2$   \\ \hline
     W-II   &   $B_{\textrm{W1}}=5,~B_{\textrm{W2}}=5$ &    $B_{\textrm{N1}}=2,B_{\textrm{N2}}=1,B_{c}=2$    \\  \hline
\end{tabular}
\label{tab:fo_wide}
\end{table}

\begin{table}
\caption{{Wideband configurations for narrowband and wideband quantizers}}
\centering
\begin{tabular}{|c|l|c|l|}
  \hline
 & Narrowband quantizer &    & Wideband quantizer \\ \hline
       N-1  & 1 codeword / 75 tones  &W-1 & $L=4, R=2$\\ \hline
      N-2  & 1 codeword / 600 tones  & W-2& $L=1, R=9$ \\ \hline
\end{tabular}
\label{tab:wide_con}
\end{table}

 We next evaluate the normalized beamforming gain of the wideband quantizer {according to}
\begin{align*}
\mathrm{G}_{\textrm{W}}\doteq \mathrm{E}_{\bh[w] } \big[ \big|  {\bff_{\ell,r}^H \bh[w]}/{\|\bh[w]  \|_2} \big|^2 \big],
\end{align*}
where $\bff_{\ell,r}$ is the chosen codeword in (\ref{eq_J4:sel_dual}). In Figs. \ref{fig_J4:wide_a} and \ref{fig_J4:wide_b},  {the normalized beamforming gains of wideband quantizer are compared with those of the narrowband quantizers.} In the legend, the first alphabet denotes the type of quantizer, the second alphabet denotes the {feedback-bit allocation} scenario in Table \ref{tab:fo_wide}, and the final digit represents the wideband configuration\footnote{In the LTE setup of scheme W-3, the first $8$ narrowband RBs have $72$ tone CSIs  and the ninth narrowband RB has $24$ tone CSIs.} in Table \ref{tab:wide_con}. The total feedback overhead of the proposed wideband quantizer is defined as
\begin{align*}
B_{\textrm{W}}&=2(B_{\textrm{W1}}+B_{\textrm{W2}})L+(2B_{\textrm{N1}}+1)RL.
\end{align*}
Numerical results verify that  wideband quantizers outperforms narrowband quantizers because {it exploits} correlation between frequency tone CSIs. The wideband quantizers also reduce feedback overhead because {they} can maintain quantization performance with less overhead compared to  narrowband quantizers.

\section{Conclusion}
{In this paper, advanced CSI quantizers based on the KP codebook structure are proposed for FD-MIMO systems using UPAs.  In the proposed quantizer designs,} we focused on detecting and quantizing a limited number of dominant {2D beams in 3D channel vectors} by exploiting DFT codebooks. The codebook for combiners was designed to {cophase and scale the quantized 2D DFT beams}. Furthermore, we analytically derived a design guideline for practical quantizers, {which is based on FD-MIMO systems with predefined {feedback-bit allocation} scenarios.}

We then developed CSI quantizers by taking the predefined feedback scenarios into account. First, a narrowband quantizer was proposed to quantize and/or combine one or two dominant {2D DFT beams}. To detect and quantize beams properly, we also developed a  multi-round beam search approach that scans both vertical and horizontal domains jointly under the moderate computational complexity. To reduce total feedback overhead, we also proposed a  wideband quantizer that utilizes the correlated information between multiple frequency tones.  {Numerical simulations verified that the proposed narrowband quantizer gives better quantization performance than previous CSI quantization techniques, and the proposed wideband quantizer further improves the quantization performance with less feedback overhead compared to the narrowband quantizer in wideband settings.}

\appendices

\section{Correlation between array response vector and  DFT codeword}
\label{sec_J4:dm}

We discuss the correlation  between the array response vector in the domain $a$ and the selected DFT codeword to quantify the quantization performance of DFT codebooks {by evaluating}
\begin{align}
\nonumber
\Gamma_{na}^2 &\doteq  \mathrm{E}\big[ \big| \bd^H_{M_a}(\psi_{n}^{a})\bc_{n}^{a} \big|^2 \big]
\\
\nonumber
&=\mathrm{E}\Big[ \max_{q \in \{ 1,\cdots,Q_n\}} \big| \bd^H_{M_a}(\psi_{n}^{a}) \ba_{M_a}(q/Q_n) \big|^2 \Big]
\\
\nonumber
&=  \frac{1}{M_a^2}\mathrm{E}\bigg[ \max_{q \in \{ 1,\cdots,Q_n\}} \bigg| \sum_{m=0}^{M_a-1} e^{-j \pi m (\psi_{n}^{a} -  {2q}/{Q_n}+1)} \bigg|^2  \bigg]
\\
\nonumber
& \stackrel{(a)} =  \frac{1}{M_a^2} \mathrm{E}\bigg[ \bigg| \sum_{m=0}^{M_a-1} e^{-j \pi m (\psi_{n}^{a} -{2{q_n}}/{Q_n}+1 )} \bigg|^2  \bigg],
\end{align}
where $(a)$ is rewritten with the  index {of} selected codeword
\begin{align*}
q_n=\argmin_{q=\{1,\cdots,Q_n\}} \big| \psi_{n}^{a} - {2q}/{Q_n}+1 \big|.
\end{align*}
The expectation over $\psi_{n}^{a} \sim \mathrm{U}(-1,1)$ is rewritten by defining the new random variable as $\psi \doteq \psi_{n}^{a} -  {2q_n}/{Q_n} +1$, which follows $\mathrm{U}( - {1}/{Q_n}, {1}/{Q_n})$ because $| \psi_{n}^{a} - {2q_n}/{Q_n}  +1 | \leq {1}/{Q_n}$. The correlation formula\footnote{{We assume that the number of {oversampled DFT codewords} is larger than the number of antennas, e.g., $Q_n \gg M_a$, in each domain $a \in \{ h,v\}$. Assuming $Q_a \gg M_a/2$, the correlation formula is always positive, i.e., $\Gamma_{na}^2>0$,  because $|\psi|<2/M_a$.}} is then computed over $\psi$, as}
\begin{align}
\nonumber
&\frac{1}{M_a^2}  \mathrm{E} \bigg[ \bigg| \sum_{m=0}^{M_a-1}   e^{-j\pi m \psi} \bigg|^2\bigg]=\frac{1}{M_a^2}  \mathrm{E} \bigg[ \sum_{\ell=0}^{M_a-1}  \sum_{m=0}^{M_a-1} e^{-j\pi(m-\ell)\psi}\bigg]
\\
\nonumber
&=\frac{1}{M_a^2}\bigg(M_a+  \sum_{\ell=0}^{M_a-1} \sum_{m > \ell}^{M_a-1} \int_{-\frac{1}{Q_n}}^{\frac{1}{Q_n}} \frac{2\cos\big(\pi(m-\ell)\psi\big)}{2/Q_n} d\psi  \bigg)
\\
\nonumber
&=\frac{1}{M_a^2}\bigg(M_a+ \sum_{\ell=0}^{M_a-1} \sum_{m > \ell}^{M_a-1} \frac{2\sin{\big(\pi (m-\ell) /Q_n \big)}}{\pi (m-\ell) /Q_n}  \bigg)
\\
\nonumber
&=\frac{1}{M_a^2}\bigg(M_a+ \sum_{q=1}^{M_a-1} \frac{2(M_a-q)\sin{\big(\pi q /Q_n\big)}}{\pi q /Q_n}  \bigg).
\end{align}



\section{Lower bound of normalized  {beamforming gain}}
\label{sec_J4:bq}

\begin{remark}
\label{rm_J4:02}
{To simplify analysis, we consider the  first order Taylor expansion of the bivariate variables, which is derived as in \cite{Ref_Ken98},
\begin{align*}
\frac{A}{B} \simeq \frac{\mathrm{E}[A]}{\mathrm{E}[B]}+\frac{\partial\big(\frac{\mathrm{E}[A]}{\mathrm{E}[B]}\big)}{\partial A}   (A-\mathrm{E}[A])+\frac{\partial\big(\frac{\mathrm{E}[A]}{\mathrm{E}[B]}\big)}{\partial B} (B-\mathrm{E}[B]).
\end{align*}
Expectation of the bivariate variables is then approximated as}
\begin{align*}
\mathrm{E} \bigg[ \frac{A}{B} \bigg] &\simeq \mathrm{E} \bigg[ \frac{\mathrm{E} [{A}]}{\mathrm{E} [{B}]} \bigg]=\frac{\mathrm{E}[A]}{\mathrm{E}[B]}.
\end{align*}
\end{remark}

The  {beamforming gain} $\mathrm{G}^{\mathrm{bq}}$ in Lemma \ref{lm_J4:bq} is  lower bounded as
\begin{align}
\nonumber
\mathrm{G}^{\mathrm{bq}} &= \mathrm{E} { \bigg[ \max_{\tilde{\bz} \in \mathbb{C}^N }\frac{|{\bh}^H\bC\tilde{\bz} |^2}{\| \bC\tilde{\bz}\|_2^2} \bigg]}
\\
\nonumber
&\stackrel{(a)}  \geq  \mathrm{E} { \bigg[\frac{\max_{\tilde{\bz} \in \mathbb{C}^N }  {|{\bh}^H\bC \tilde{\bz}|^2}}{\| \bC\|_2^2 }   \bigg]}
\\
\nonumber
& \stackrel{(b)} = \mathrm{E} { \bigg[  \frac{\|{\bh}^H\bC\|_2^2}{\| \bC\|_2^2  } \bigg]}
\\
\label{eq_J4:lb_mono}
& {\stackrel{(c)} \simeq  \frac{\mathrm{E}\big[\|{\bh}^H\bC\|_2^2\big]}{\mathrm{E}\big[\|\bC\|_2^2\big]},}
\end{align}
{where the inequality in  $(a)$ is  based on  $\|\bC\tilde{\bz} \|_2^2 \leq \|\bC \|_2^2 \|\tilde{\bz} \|_2^2$ and $\|\tilde{\bz} \|_2^2=1$, $(b)$ holds when $\tilde{\bz} \doteq {\bC^H\bh}/\| {\bC^H\bh} \|_2$}, and $(c)$ is derived based on Remark \ref{rm_J4:02}.

To complete the lower bound in (\ref{eq_J4:lb_mono}), we first compute the expectation of {two-norm squared} of  the effective  channel vector
\begin{align}
\nonumber
\mathrm{E} \big[  \| {{\bh}}^H\bC\|_2^2 \big] &= \mathrm{E} \Big[   \big\| {{\bh}}^H[ \bc_{1}^{v} \otimes \bc_{1}^{h},\cdots, \bc_{N}^{v} \otimes \bc_{N}^{h} ]  \big\|_2^2 \Big]
\\
\nonumber
&= \sum_{n=1}^N \mathrm{E} \Big[ \big| {{\bh}}^H(\bc_{n}^{v}\otimes \bc_{n}^{h}) \big|^2\Big]
\\
\label{eq_J4:norm_ec}
& \stackrel{{(a)}} =  \frac{1}{M} \bigg( PN+ \sum_{n=1}^N\sum_{q=n}^{P}\frac{M\Gamma_{nv}^2 \Gamma_{nh}^2 -1}{q}\bigg),
\end{align}
where $(a)$ is derived by using the correlation between the channel vector and the $n$-th selected DFT codeword $\mathrm{E} \big[ | {{\bh}}^H(\bc_{n}^{v}\otimes \bc_{n}^{h}) |^2  \big]$ that will be discussed in Appendix \ref{sec_J4:cc}.

\begin{figure*}[!t]
\setcounter{equation}{36}
\begin{align}
\nonumber
\mathrm{E} [(\bc_{c}^{v} \otimes  \bc_{c}^{h})^H {{\bh}} {{\bh}}^H (\bc_{d}^{v} \otimes  \bc_{d}^{h}) ]  & = \bigg(\sum_{p=1}^P \alpha_p^* (\bd^H_{M_v}(\psi_{p}^{v} )\bc_{c}^{v} )( \bd^H_{M_v}(\psi_{p}^{h})\bc_{c}^{h} ) \bigg)^* \bigg(\sum_{q=1}^P \alpha_q^* (\bd^H_{M_v}(\psi_{q}^{v} )\bc_{d}^{v} )( \bd^H_{M_v}(\psi_{q}^{h})\bc_{d}^{h} ) \bigg)
 \\
 \label{eq_J4:ent}
 &= \sum_{p=1}^P \mathrm{E} \big[  |\alpha_p|^2 \big]\mathrm{E} \big[  (\bc_{c}^v)^H\bd_{M_v}({\psi_{p}^{v}}) \bd_{M_v}^H({\psi_{p}^{v}}) \bc_{d}^{v} \big] \mathrm{E} \big[  (\bc_{c}^h)^H\bd_{M_v}({\psi_{p}^{h}}) \bd^H_{M_v}({\psi_{p}^{h}}) \bc_{d}^{h} \big].
\end{align}
\hrulefill
\setcounter{equation}{32}
\end{figure*}

We next consider the  set of $N$ DFT vectors $\bC$ in (\ref{eq_J4:C}). The expectation of {two-norm squared} of $\bC$   is approximated as
\begin{align}
\nonumber
\mathrm{E} \big[ \| \bC \|_2^2 \big] &=   \mathrm{E} \bigg[ \max_{\tilde{\bx} \in \mathbb{C}^N} {\tilde{\bx}^H (\bC^H \bC)\tilde{\bx} } \bigg]
 \\
 \label{eq_J4:Csqu}
 & \simeq  \max_{\tilde{\bx} \in \mathbb{C}^N} {\tilde{\bx}^H \mathrm{E}\big[\bC^H \bC \big] \tilde{\bx} }
 \end{align}
subject to $\| \tilde{\bx} \|_2^2=1$. Note that $\mathrm{E}\big[\bC^H \bC \big]$ is computed as
\begin{align}
\nonumber
& \mathrm{E} \left[\begin{array}{ccc}(\bc_{1}^{v})^H\bc_{1}^{v} \otimes (\bc_{1}^{h})^H\bc_{1}^{h} &  \cdots & (\bc_{1}^{v})^H\bc_{n}^{v}\otimes (\bc_{1}^{h})^H\bc_{n}^{h}
 \\
\vdots &    \ddots  &     \vdots
 \\
(\bc_{n}^{v})^H\bc_{1}^{v} \otimes (\bc_{n}^{h})^H\bc_{1}^{h}&  \cdots    &     (\bc_{n}^{v})^H\bc_{n}^{v}\otimes (\bc_{n}^{h})^H\bc_{n}^{h}
 \\
\end{array} \right]
\\
\label{eq_J4:cor_Q}
& = \left[\begin{array}{ccc} 1 &  \cdots & \frac{1}{M_v M_h}
 \\
\vdots &      \ddots   &   \vdots
 \\
\frac{1}{M_v M_h} &  \cdots     &  1
 \\
\end{array} \right]
\end{align}
{where $\bc_{n}^{a}=\ba_{M_a}(q_n/Q_n)$ with $Q_n=2^{B_n}$ is  chosen as in (\ref{eq_J4:sel_dft}).  {Because we assume that beam directions are uniformly distributed $\psi_{n}^{a}\sim \mathrm{U}(-1,1)$, $\bc_{n}^{a}$ can be chosen as one of $Q_n=2^{B_n}$ codewords in the DFT codebook $\mathcal{A}_{B_n}^{a}$ with equal probabilities. For this reason, we can obtain $\mathrm{E} \big[ (\bc_{c}^{a})^H\bc_{d}^{a} \big]$ in (\ref{eq_J4:cor_Q})  by computing its arithmetic mean of the {beamforming gain} between two different codewords ($c\ne d$)  as}
\begin{align}
\nonumber
&\mathrm{E} \big[ (\bc_{c}^{a})^H\bc_{d}^{a} \big] =  \frac{1}{Q_c Q_d} \sum_{q_c=1}^{Q_c} \sum_{q_d=1}^{Q_d} \ba_{M_a}^H(q_c/Q_c)\ba_{M_a}(q_d/Q_d) 
\\
\nonumber
&= \frac{1}{M_aQ_c Q_d} \Bigg[\sum_{q_c=1}^{Q_c} \sum_{q_d=1}^{Q_d}  1  + \sum_{m=1}^{M_a-1}  \sum_{q_c=1}^{Q_c} \sum_{q_d=1}^{Q_d}  e^{ \frac{j 2\pi  m q_d}{Q_d}} e^{ \frac{-j 2\pi m q_c }{Q_c}} \Bigg]
\\
\nonumber
&= \frac{1}{M_aQ^2} \Bigg[ Q^2  + \sum_{m=1}^{M_a-1}  \bigg( \frac{1-e^{{j2 \pi m}}}{1-e^{\frac{j2 \pi m}{Q_d}}}\bigg) \bigg( \frac{1-e^{-j 2 \pi m}}{1-e^{-\frac{j2 \pi m}{Q_c}}} \bigg)  \Bigg]
\\
\label{eq_J4:ari}
&= \frac{1}{M_a}.
\end{align}

{Based on $\mathrm{E}\big[\bC^H \bC \big]$, the expectation of {two-norm squared} of $\bC$ in (\ref{eq_J4:Csqu}) is approximated as
\begin{align}
\nonumber
 \mathrm{E} \big[ \| \bC \|_2^2 \big]    & \simeq   \mathfrak{eig}_{1} \big\{  \mathrm{E}\big[\bC^H \bC \big] \big\}
 \\
 \nonumber
 &=\frac{\mathfrak{eig}_{1} \{ \mathbf{1}_{N,N} \}}{M} + \frac{M-1}{M}
 \\
  \label{eq_J4:cor_F}
 &\stackrel{(a)}=\frac{M+N-1}{M},
\end{align}
where $(a)$ is derived because $\mathfrak{eig}_{1} \{ \mathbf{1}_{N,N} \}=N$, which holds when $\tilde{\bx}=\frac{1}{\sqrt{N}}[1,\cdots,1]^T \in \mathbb{C}^N$.}

Finally, the lower bound of $ \mathrm{G}^{\mathrm{bq}}$ in (\ref{eq_J4:lb_mono}) is approximated by plugging the  derived formulas in   (\ref{eq_J4:norm_ec}) and (\ref{eq_J4:cor_F}) into (\ref{eq_J4:lb_mono})  as
\begin{align*}
 \mathrm{G}^{\mathrm{bq}} \simeq  \frac{P}{M+N-1}\bigg(N+ \sum_{n=1}^N \sum_{q=n}^P\frac{M \Gamma^2_{nv}\Gamma^2_{nh}-1}{qP}\bigg).
\end{align*}


\section{Covariance matrix of effective channel vector}
\label{sec_J4:cm}

{Each entry of $\bR$ in  (\ref{eq_J4:cov}) is rewritten  in (\ref{eq_J4:ent}).} Note that $\mathrm{E} \big[  (\bc_{c}^{a})^H\bd_{M_a}({\psi_{p}^{a}}) \bd_{M_a}^H({\psi_{p}^{a}}) \bc_{d}^{a} \big]$ in (\ref{eq_J4:ent}) is computed depending on the different cases as follows:

\textbf{Case 1:} {$p=c=d$}.
\begin{align*}
\mathrm{E} \big[  (\bc_{c}^{a})^H\bd_{M_a}({\psi_{p}^{a}}) \bd_{M_a}^H({\psi_{p}^{a}}) \bc_{d}^{a} \big]
= \Gamma^2_{ac},
\end{align*}
{where $\Gamma_{ac}^2$ is derived in Appendix \ref{sec_J4:dm}.}

\textbf{Case 2:}  {$p\ne c,~c=d$}.
\begin{align*}
\mathrm{E} \big[  (\bc_{c}^{a})^H\bd_{M_a}({\psi_{p}^{a}}) \bd_{M_a}^H({\psi_{p}^{a}}) \bc_{d}^{a} \big] &= \mathrm{E} \Big[  \big|\bd^H_{M_a}({\psi_{p}^{a}}) \ba_{M_a}(q_c/Q_c) \big|^2 \Big]
\\
& \stackrel{(a)} =\frac{1}{M_a},
\end{align*}
where $q_c$ is chosen as in (\ref{eq_J4:sel_dft}). Note that $(a)$ is derived  by computing  the arithmetic mean.

\textbf{Case 3:} {$p=c,~c \ne d$}.
\begin{align*}
\mathrm{E} \big[  (\bc_{c}^{a})^H\bd_{M_a}({\psi_{p}^{a}}) \bd_{M_a}^H({\psi_{p}^{a}}) \bc_{d}^{a} \big]& =  {\Gamma_{ac}}  \mathrm{E} \big[   \bd^H_{M_a}({\psi_{c}^{a}}) \bc_{d}^{a} \big]
\\
&  \stackrel{(a)} = \frac{{\Gamma_{ac}}}{M_a},
\end{align*}
where $(a)$ is derived because
\begin{align*}
\mathrm{E} \big[   \bd^H_{M_a}({\psi_{c}^{a}}) \bc_{d}^{a} \big]& =\mathrm{E} \big[   \bd^H_{M_a}({\psi_{c}^{a}}) \ba_{M_a}(q_d/Q_d) \big]
\\
&= \frac{1}{M_a}\sum_{m=0}^{M_a-1} \mathrm{E} \Big[ e^{-j\pi m (\psi_{c}^{a}-\frac{2q_d}{Q_d}+1) } \Big]
\\
&  \stackrel{(b)}=  \frac{1}{M_a} \sum_{m=0}^{M_a-1} \int_{-1}^{1} \frac{e^{-j\pi m \phi}}{2} d\phi
\\
&=\frac{1}{M_a}.
\end{align*}
Note that  $(b)$ is derived because $\phi, \psi_{c}^{a} \sim \mathrm{U}(-1,1)$ with the definition of $\phi \doteq \psi_{c}^{a}-\frac{2q_d}{Q_d}+1$ for any $q_d \in \{1,\cdots,Q_d \}$ in (\ref{eq_J4:sel_dft}).

\textbf{Case 4:} $p \ne c,~p  \ne d,~c \ne d$.
\begin{align*}
&\mathrm{E} \big[  (\bc_{c}^{a})^H\bd_{M_a}({\psi_{p}^{a}}) \bd_{M_a}^H({\psi_{p}^{a}}) \bc_{d}^{a} \big]
\\
&= \frac{1}{M_a^2} \mathrm{E} \bigg[  \sum_{\ell=0}^{M_a-1}e^{j\pi \ell(\psi_{p}^{a}-\frac{2q_c}{Q_c}+1)} \sum_{m=0}^{M_a-1}e^{-j\pi m(\psi_{p}^{a}-\frac{2q_d}{Q_d}+1)}       \bigg]
\\
&  =  \frac{1}{M_a^2} \sum_{\ell=0}^{M_a-1}\sum_{m=0}^{M_a-1} e^{j\pi(\ell-m)}
\\
&~~~~~~~~~~~~~~~~~~~~~~~~~\mathrm{E} \Big[ e^{j2\pi( \frac{m q_d}{Q_d}- \frac{\ell q_c}{Q_c})} \Big] \int_{-1}^1 \frac{e^{j\pi(\ell-m)\psi_{p}^{a}}}{2} d\psi_{p}^{a}
\\
& =  \frac{1}{M_a^2} \sum_{m=0}^{M_a-1}   \mathrm{E} \Big[ e^{j2\pi m( \frac{ q_d}{Q_d}- \frac{ q_c}{Q_c})} \Big]
\\
& \stackrel{(a)} =\frac{1}{M_a^2},
\end{align*}
where $(a)$ is  derived by computing the arithmetic mean.

\section{Quantization performance of  $\mathcal{Z}_{B_{c}}$}
\label{sec_J4:bc}

{The normalized {beamforming gain} between  the effective channel vector and the selected combiner is lower bounded as
\begin{align}
\setcounter{equation}{37}
\nonumber
\mathrm{G}^{\mathrm{bc}} &=\mathrm{E} \bigg[  \frac{| \bee_{\hat{u}}^H\bR\bw|^2}{(\bw^H\bR\bw)(\bee_{\hat{u}}^H \bR \bee_{\hat{u}}) } \bigg]
\\
\nonumber
&\stackrel{(a)}      =\mathrm{E}  \bigg[  \frac{| \bee_{\hat{u}}^H\bR(\bee_{\hat{u}} \cos\theta_{\hat{u}} + \bk_{\hat{u}} \sin \theta_{\hat{u}}) |^2}{(\bw^H\bR\bw)(\bee_{\hat{u}}^H \bR \bee_{\hat{u}})  } \bigg]
\\
\nonumber
&  =   \mathrm{E}  \Bigg[ \frac{\bee_{\hat{u}}^H\bR\bee_{\hat{u}}}{\bw^H\bR\bw} \bigg|1+\frac{|\bee_{\hat{u}}^H\bR\bk_{\hat{u}}|}{|\bee_{\hat{u}}^H\bR\bee_{\hat{u}}|} \tan\theta_{\hat{u}} e^{j \angle\bee_{\hat{u}}^H\bR\bk_{\hat{u}}  } \bigg|^2\cos^2\theta_{\hat{u}} \Bigg]
\\
\nonumber
&\stackrel{(b)} \simeq       \mathrm{E}  \bigg[   \frac{\bee_{\hat{u}}^H\bR\bee_{\hat{u}} \cos^2\theta_{\hat{u}} }{\bw^H\bR\bw}    \bigg]
\\
\label{eq_J4:bc}
&\stackrel{(c)} \simeq \frac{\mathrm{E} [    {\bee_{\hat{u}}^H\bR\bee_{\hat{u}}}   \cos^2\theta_{\hat{u}}]} {\mathrm{E} [\bw^H\bR\bw]}
%
\end{align}
In (\ref{eq_J4:bc}), $(a)$ is based on $\bw  \doteq \bee_{\hat{u}} \cos \theta_{\hat{u}} + \bk_{\hat{u}} \sin \theta_{\hat{u}}$ with  $\bee_{\hat{u}} \perp \bk_{\hat{u}}$,  $(b)$ is approximated because $|\bee_{\hat{u}}^H\bR\bk_{\hat{u}}| \ll  |\bee_{\hat{u}}^H\bR\bee_{\hat{u}}|$ and $\tan \theta_{\hat{u}} \ll 1$, and  $(c)$ is approximated based on Remark \ref{rm_J4:02} in Appendix \ref{sec_J4:bq}.}


{Although  $\mathrm{G}^{\mathrm{bc}}$ is  simplified in (\ref{eq_J4:bc}), it is  still difficult to solve in most cases. In the special case of $N=2$, the equal gain vectors can be defined as
\begin{align*}
\bw \doteq \frac{e^{j\nu}}{\sqrt{2}}[1,~e^{j\upsilon}]^T,~~~\bee_{u} \doteq\frac{1}{\sqrt{2}}[1,~e^{j\frac{2\pi u}{U}}]^T
\end{align*}
using $\nu,\upsilon \sim \mathrm{U}(-\pi,\pi)$, and the {beamforming gain} in (\ref{eq_J4:bc}) is then derived such as
\begin{align*}
\frac{\mathrm{E} [    {\bee_{\hat{u}}^H\bR\bee_{\hat{u}}}   \cos^2\theta_{\hat{u}}]} {\mathrm{E} [\bw^H\bR\bw]} & \stackrel{(a)}=  \frac{\mathrm{E} [    {\bee_{\hat{u}}^H\bR\bee_{\hat{u}}} (1+\cos \hat{\nu})]} {2\mathrm{E} [\bw^H\bR\bw]}
\\
&\stackrel{(b)}=  \frac{   \sum_{\ell=1}^U \mathrm{E} \big[    {\bee_{\hat{u}}^H\bR\bee_{\hat{u}}} (1+\cos \hat{\nu})~|~\hat{u}=\ell \big]} {2U \mathrm{E} [\bw^H\bR\bw]}
\\
& =  \frac{  \sum_{\ell=1}^U       \frac{U}{2 \pi}\int_{-\frac{\pi}{U}}^{\frac{\pi}{U}} {\bee_{\ell}^H\bR\bee_{\ell}} (1+\cos \tau) d\tau} {2U\mathrm{E} [\bw^H\bR\bw]}
\\
&=  \frac{ \frac{1}{U} \sum_{\ell=1}^U     {\bee_{\ell}^H\bR\bee_{\ell}} } {2\mathrm{E} [\bw^H\bR\bw]}\Big(1 +\frac{U}{\pi} \sin \frac{\pi}{U} \Big)
\\
&\stackrel{(c)}=  \frac{1}{2}\Big(1 +\frac{U}{\pi} \sin \frac{\pi}{U} \Big).
\end{align*}
Note that $(a)$ is derived based on the definition $\hat{\upsilon} \doteq \upsilon - \frac{2\pi\hat{u}}{U}$ that follows $\mathrm{U}(-\frac{\pi}{U},\frac{\pi}{U})$ because $| \upsilon - \frac{2\pi\hat{u}}{U} | \leq \frac{\pi}{U}$, $\upsilon \sim \mathrm{U}(-\pi,\pi)$, and  $\hat{u} = \argmin\limits_{u \in \{1,\cdots,U\} } | \upsilon - \frac{2\pi u}{U} |$ based on Assumption \ref{am:bc}. In addition, $(b)$ is derived by computing its arithmetic mean because $\hat{u}$ is equally probable from $1$ to $U$, and $(c)$ is derived because
\begin{align*}
\mathrm{E}[{ \bw^H\bR \bw }]&=({\bR_{1,1}+\bR_{2,2}})/{2}+|\bR_{1,2}|\mathrm{E} [\cos(\upsilon+\angle \bR_{1,2}) ]
\\
&=({\bR_{1,1}+\bR_{2,2}})/{2},
\end{align*}
and the arithmetic mean of $  {\bee_{\ell}^H\bR\bee_{\ell}}$ is derived as}
\begin{align*}
 \sum_{\ell=1}^U \frac{   {\bee_{\ell}^H\bR\bee_{\ell}}}{U} &=\frac{\bR_{1,1}+\bR_{2,2}}{2}+\sum_{\ell=1}^U \frac{|\bR_{1,2}| \cos\big(\frac{2 \pi {\ell}}{U}+\angle \bR_{1,2}\big)}{U}
\\
&=({\bR_{1,1}+\bR_{2,2}})/{2}.
\end{align*}

\section{Correlation between channel vector and DFT codeword}
\label{sec_J4:cc}
We derive a correlation between the channel in (\ref{eq_J4:simple_channel}) and the $n$-th selected {2D DFT beam}
\begin{align}
\nonumber
 &\mathrm{E} \Big[ \big| {{\bh}}^H(\bc_{n}^{v}\otimes \bc_{n}^{h}) \big|^2 \Big]
  \\
\nonumber
  &= \mathrm{E} \bigg[ \bigg| \sum_{p=1}^{P}  \alpha_{p}^{*} \big( \bd_{M_v}^H(\psi_{p}^{v})\bc_{n}^{v}\big) \big( \bd_{M_h}^H(\psi_{p}^{h})\bc_{n}^{h} \big) \bigg|^2 \bigg]
\\
\nonumber
& \stackrel{{(a)}} =    \sum_{p=1}^{P}  \mathrm{E} \big[ | \alpha_{p} |^2 \big]  \mathrm{E} \big[ \big| \bd_{M_v}^H(\psi_{p}^{v})\bc_{n}^{v}\big|^2 \big] \mathrm{E} \Big[ \big| \bd_{M_h}^H(\psi_{p}^{h})\bc_{n}^{h} \big|^2  \Big]
\\
\nonumber
& \stackrel{{(b)}}=  \mathrm{E} \big[ | \alpha_{n} |^2 \big]\Gamma_{nv}^2\Gamma_{nh}^2+\sum_{p \ne n}^P \frac{\mathrm{E} \big[ | \alpha_{p}|^2 \big]}{M_v M_h}
\\
\label{eq_J4:00}
& =  \mathrm{E} \big[ | \alpha_{n} |^2 \big]\Gamma_{nv}^2\Gamma_{nh}^2+  \frac{P- \mathrm{E} \big[ | \alpha_{n} |^2 \big]}{M},
\end{align}
where $\Gamma_{na}^2 \doteq \mathrm{E}{\big[ \big|  \bd^H_{M_a}(\psi_{n}^{a}) \bc_{n}^{a}  \big|^2 \big]}$ is derived in Appendix \ref{sec_J4:dm}. Note that ${(a)}$ is derived because $\mathrm{E}[\alpha_p^*\alpha_q]=0$ when $p \ne q$ and $ {(b)}$ is derived because
\begin{align*}
&\mathrm{E}\Big[\big| \bd_{M_a}^H(\psi_{p}^{a})\bc_{n}^{a} \big|^2\Big]=\mathrm{E}\Big[\big| \bd_{M_a}^H(\psi_{p}^{a})\ba_{M_a}(q_n/Q_n) \big|^2\Big]
\\
&= \frac{1}{M_a^2} \mathrm{E} \bigg[  \sum_{\ell=0}^{M_a-1}e^{j\pi\ell(\psi_{p}^{a}-\frac{2q_n}{Q_n}+1)} \sum_{m=0}^{M_a-1}e^{-j\pi m(\psi_{p}^{a}-\frac{2q_n}{Q_n}+1) }       \bigg]
\\
&  =  \frac{1}{M_a^2} \sum_{\ell=0}^{M_a-1}\sum_{m=0}^{M_a-1} \mathrm{E} \big[ e^{j \pi(m -\ell) (\frac{2q_n}{Q_n}-1)} \big] \int_{-1}^1 \frac{e^{j\pi(\ell-m)\psi_{p}^{a}}}{2} d\psi_{p}^{a}
\\
&  =  \frac{1}{M_a^2} \sum_{m=0}^{M_a-1} \sum_{m=\ell}^{M_a-1} \mathrm{E} \big[ e^{j \pi(m -\ell) (\frac{2q_n}{Q_n}-1)}  \big] \frac{\sin(\pi(\ell-m))}{\pi(\ell-m)}
\\
&=\frac{1}{M_a}.
\end{align*}


To complete the formula in (\ref{eq_J4:00}), we compute the power of the $n$-th largest channel gain  $\mathrm{E} [ | \alpha_{n} |^2 ]$. Without loss of generality, we assume that the magnitude of channel gains are in descending order, i.e., $|\alpha_1|^2\geq \cdots \geq |\alpha_P|^2$. The channel gain $\mathrm{A} \doteq |\alpha_p|^2$ follows $\chi_2^2$ that is characterized by the cumulative distribution function (cdf) of
\begin{align*}
\mathrm{F}_{\mathrm{A}}(a)=1-e^{-a}
\end{align*}
because $\alpha_p \sim \mathcal{CN}(0,1)$. {We now consider the  $k$-th order statistic ($k$-th smallest order statistic) of $P$ i.i.d exponentially distributed random variables $\mathrm{A}_k \doteq  |\alpha_{P-k+1}|^2$. Then, we refer to \cite{Ref_Dav80} for defining the pdf of $\mathrm{A}_k$, yielding}
\begin{align*}
f_{\mathrm{A}_k}(a)&= P \binom{P-1}{k-1} (1-e^{-a})^{k-1}e^{-a(P-k+1)}
\\
&\stackrel{(a)}=P \binom{P-1}{k-1} \sum_{q=0}^{k-1}\binom{k-1}{q}(-1)^q e^{-a(q+P-k+1)},
\end{align*}
where $(a)$ is derived based on the binomial expansion formula. Then, the expectation of $k$-th order statistic is defined as
\begin{align*}
&\mathrm{E}\big[\mathrm{A}_k\big] 
=  \sum_{q=0}^{k-1} \frac{P! \binom{k-1}{q}(-1)^q}{(P-k)! (k-1)!} \int_{0}^{\infty} a e^{-a(q+(P-k+1))} da
\\
&=  \sum_{q=P-k+1}^{P} \frac{P! (-1)^{q-(P-k+1)}  }{(P-q)!q!} \bigg[ \frac{ (q-1)\cdots(q-(P-k)) }{q (P-k)!  } \bigg]
\\
& \stackrel{(a)}= \sum_{q=1}^{P} \frac{\binom{P}{q}(-1)^{q-(P-k+1)} }{q}\bigg[ \frac{q-1}{1} \frac{q-2}{2} \cdots \frac{q-(P-k)}{P-k}   \bigg]
\\
&\stackrel{(b)}=   \Bigg[ \sum_{q=1}^{P} \binom{P}{q}(-1)^{q}\Bigg] \sum_{\ell=1}^{P-k}\frac{1}{\ell} + \sum_{q=1}^{P}\frac{\binom{P}{q}(-1)^{q+1}}{q}
\\
&\stackrel{(c)}  =  \sum_{\ell=1}^{P-k}\frac{1}{\ell} \Bigg[\sum_{q=0}^{P} \binom{P}{q}(-1)^{q} - \binom{P}{0}(-1)^0  \Bigg] + \sum_{q=1}^{P}\frac{1}{q}
\\
&\stackrel{(d)}  =-\sum_{\ell=1}^{P-k}\frac{1}{\ell}  + \sum_{q=1}^{P}\frac{1}{q}
\\
&= \sum_{q=P-k+1}^{P}\frac{1}{q}.
\end{align*}
Notice that $(a)$ is derived {because $\sum_{q=b}^P f(q)(q-b)=\sum_{q=b+1}^P f(q)(q-b)+f(b)(b-b)$
for any function $f( \cdot )$}, $(b)$ is derived because $\sum_{q=1}^{P}\frac{\binom{P}{q}(-1)^{q+1}}{q}=\sum_{q=1}^{P}\frac{\binom{P}{q}(-1)^{q-1}}{q}$, $(c)$ is derived based on $\sum_{k=1}^n\frac{\binom{n}{k}(-1)^{k+1}}{k}=\sum_{k=1}^n\frac{1}{k}$, and $(d)$ is derived based on $\sum_{k=0}^n (-1)^k\binom{n}{k}=0$.
{We now compute the $n$-th largest channel gain as}
\begin{align*}
\mathrm{E}\big[|\alpha_n|^2\big]&=\mathrm{E}\big[\mathrm{A}_{P-n+1}\big]=\sum_{q=n}^{P}\frac{1}{q}.
\end{align*}

Finally, the correlation coefficient is rewritten by plugging $|\alpha_n|^2$ into (\ref{eq_J4:00}).
\begin{align*}
\mathrm{E} \big[ \big| {{\bh}}^H(\bc_{n}^{v}\otimes \bc_{n}^{h}) \big|^2 \big] 
&  =  \frac{1}{M} \bigg(   P +         \sum_{q=n}^P \frac{M\Gamma_{nv}^2 \Gamma_{nh}^2-1}{q} \bigg).
\end{align*}

\bibliographystyle{IEEEtran}
\bibliography{refs_jiho}

\end{document}